\documentclass[10pt, conference, letterpaper]{IEEEtran}

\IEEEoverridecommandlockouts              

\usepackage{fancyhdr}
\usepackage{graphicx}
\usepackage[utf8]{inputenc}

\graphicspath{{figures/}}

\usepackage{enumitem}
\usepackage{url}
\usepackage{wrapfig}
\usepackage{units}

\usepackage[noadjust]{cite} 

\usepackage[printonlyused]{acronym}

\usepackage{calc}
\usepackage{graphicx}
\usepackage[lined,boxed, vlined,ruled,linesnumbered]{algorithm2e}
\usepackage{amsmath,amssymb,amsthm,amsfonts,amsbsy}

\usepackage{algpseudocode}
\usepackage{color}
\usepackage[table]{xcolor} 

\usepackage[caption=false]{subfig}
\usepackage{dblfloatfix}
\usepackage{nicefrac}
\usepackage{tcolorbox}
\usepackage{txfonts}
\usepackage{soul}

\usepackage{multirow}

\newcolumntype{C}[1]{>{\centering\arraybackslash}p{#1}}

\usepackage{booktabs}

\definecolor{BgGray}{gray}{0.7}%
\definecolor{BgGray2}{gray}{0.96}%
\definecolor{RowColorOdd}{named}{BgGray2}%
\definecolor{RowColorEven}{named}{white}%
\definecolor{comments}{gray}{.5}
\definecolor{Gray}{gray}{0.85}

\usepackage{pifont}
\usepackage[multiuser,nomargin,marginclue,footnote]{fixme}
\fxusetheme{color}
\fxsetup{targetlayout=colorcb}
\FXRegisterAuthor{all}{anall}{ALL}
\FXRegisterAuthor{md}{anmd}{MD}
\FXRegisterAuthor{tolja}{antolja}{TOLJA}
\FXRegisterAuthor{pg}{anpg}{PG}
\FXRegisterAuthor{mch}{anmc}{MC}
\FXRegisterAuthor{cp}{ancp}{CP}
\FXRegisterAuthor{awo}{anawo}{AWO}

\fxsetup{status=draft}

\newcommand{\proposalName}{OfdmFi}

\begin{document}
\providetoggle{techreport}
\settoggle{techreport}{false}

\title{OfdmFi: Enabling Cross-Technology Communication Between LTE-U/LAA and WiFi}

\author{
\IEEEauthorblockN{Piotr Gaw{\l}owicz, Anatolij Zubow, Suzan Bayhan and Adam Wolisz}
\IEEEauthorblockA{\{gawlowicz, zubow, bayhan, wolisz\}@tkn.tu-berlin.de}
Technische Universität Berlin, Germany
}

\maketitle


\begin{abstract}

Despite exhibiting very high theoretical data rates, in practice, the performance of \mbox{LTE-U/LAA} and WiFi networks is severely limited under cross-technology coexistence scenarios in the unlicensed 5\,GHz band. 
As a remedy, recent research shows the need for collaboration and coordination among co-located networks. However, enabling such collaboration requires an information exchange that is hard to realize due to completely incompatible network protocol stacks. %
We propose OfdmFi, the first cross-technology communication scheme that enables direct bidirectional over-the-air communication between \mbox{LTE-U/LAA} and WiFi with minimal overhead to their legacy transmissions. 
Requiring neither hardware nor firmware changes in commodity technologies, OfdmFi leverages the standard-compliant possibility of generating message-bearing power patterns, similar to punched cards from the early days of computers, in the time-frequency resource grid of an OFDM transmitter which can be cross-observed and decoded by a heterogeneous OFDM receiver. 
As a proof-of-concept, we have designed and implemented a prototype using commodity devices and SDR platforms. Our comprehensive evaluation reveals that OfdmFi achieves robust bidirectional CTC between both systems with a data rate up to 84\,kbps, which is more than 125$\times$ faster than state-of-the-art.

\end{abstract}

%

\section{Introduction}

With the explosive increase in cellular traffic on one side and the proliferation of massive Internet of Things on the other~\cite{cisco2019}, unlicensed radio spectrum~(e.g., ISM/UNII) becomes crowded by numerous wireless network devices with technologies ranging from LTE, WiFi, ZigBee, and Bluetooth. 
Unfortunately, the heterogeneous technologies with diverse operation principles are largely oblivious to each other and their naive (i.e. uncoordinated) coexistence leads to severe cross-technology interference~(CTI), which is a major cause of network performance degradation~\cite{5986182, Liang}.

Operating in unlicensed bands requires LTE to coexist fairly with WiFi that was so far the dominant technology in 5\,GHz spectrum.
Although both technologies are already very advanced (i.e., their newest generations provide peak data rates in the order of 1\,Gbps), under coexistence scenarios they still rely on rather primitive coexistence schemes based on energy-sensing and hence suffer from frequent collisions and significant throughput degradation~\cite{chai2016lte}.
This impact is largely attributed to the lack of understanding of each other's waveforms as well as the differences in their operation, e.g., carrier sensing sensitivity, (a)synchronous access principles or contention window adaptation~\cite{8292751}.

Recently, it has become clear that an explicit and coordinated collaboration among co-located heterogeneous networks is needed to efficiently tackle CTI, ensure fair coexistence and bring performance breakthroughs in spectrum sharing~\cite{yin2018explicit,sc2_darpa,sagari2015coordinated,LTERadar}. 
However, explicit collaboration requires direct communication between heterogeneous devices that is hard to realize due to incompatible network protocol stacks.
Note that the coexisting networks can be coordinated by a single central controller. Unfortunately, this possibility cannot be directly utilized as even if the networks are co-located, due to different physical layers, they do not know whom they are interfering with.
Hence, in recent years we have seen a boom of wireless cross-technology communication~(CTC) designs, e.g., ~\cite{Esense,GapSense,cmorse,TransparentCTC, WeBee,gawlowicz18_infocom,emf,wizig}, that can be used for cross-technology neighbor discovery and identification~\cite{gawlowicz18_infocom}.
Furthermore, in~\cite{zubow2018practical,bayhan2018cogapsWoWMoM,xzero_demo}, we have presented a novel collaboration scheme between LTE-U and WiFi enabled by CTC, that allows LTE-U BS equipped with multiple antennas to steer a null towards WiFi stations in order to mitigate the CTI.

Despite these recent advances, prior CTC solutions pertain mostly to WiFi, ZigBee, and Bluetooth. 
So far there are only two CTC schemes addressing unlicensed LTE and WiFi case, however, both come with some shortcomings.
LtFi~\cite{gawlowicz18_infocom} provides only simplex over-the-air CTC from LTE-U eNB to WiFi AP, while the reverse direction is realized over the wired Internet inducing delay in the order of tens of milliseconds.
ULTRON~\cite{chai2016lte} is not a generic CTC scheme as it only allows embedding valid WiFi CTS frame in LTE transmissions for the purpose of cross-technology channel reservation.
\begin{figure}[t!] 
	\centering
	\vspace{-5pt}
	\includegraphics[width=1.0\linewidth]{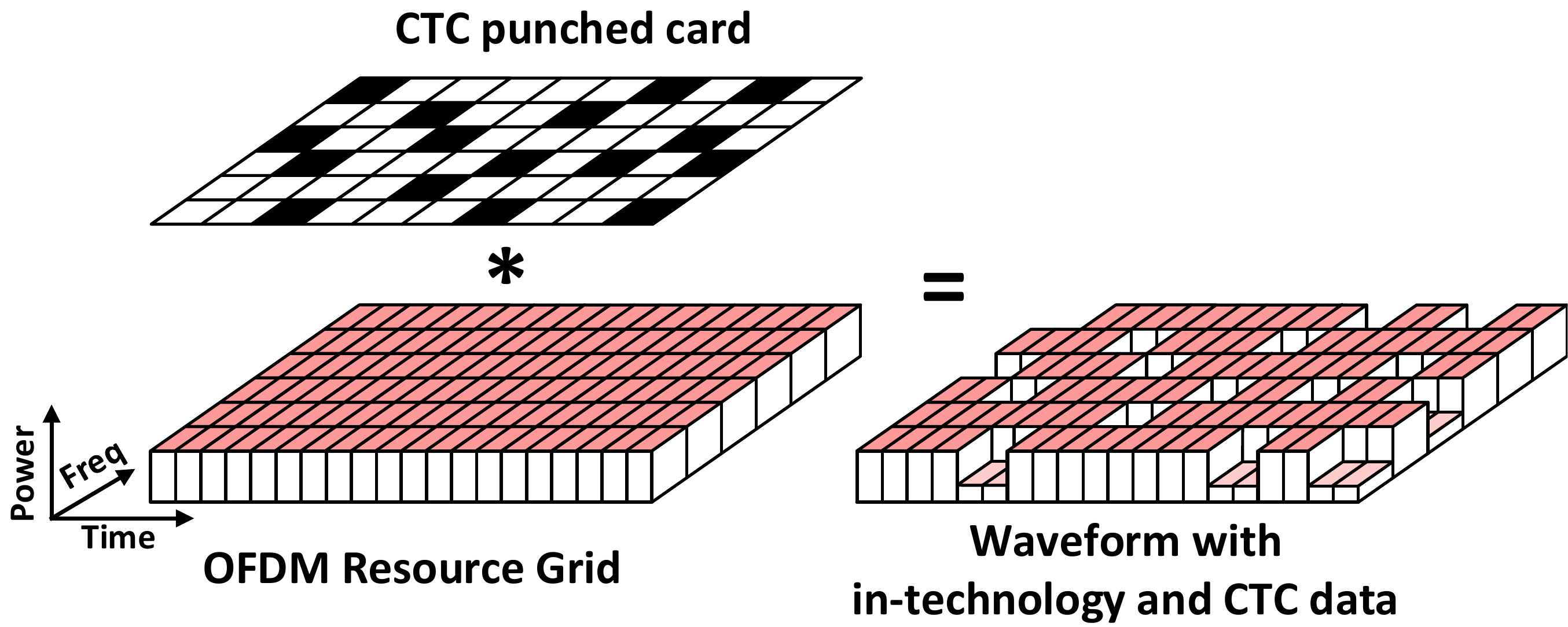}
	\vspace{-20pt}
	\caption{\proposalName~creates CTC punched cards using OFDM resources.}
	\label{fig:overlay_ctc}
	\vspace{-15pt}
\end{figure}

\smallskip
In this paper, we present \textbf{\proposalName}, a CTC scheme that enables direct over-the-air communication between LTE-U/LAA base stations and WiFi access points and hence empowers co-located wireless networks of both technologies to establish common control channel and implement coexistence strategies minimizing the impact of CTI.
\proposalName~features an innovative usage of power modulation for CTC and achieves high efficiency by building on new insights on cross-technology OFDM signal reception.
Specifically, having the possibility to create cross-observable power patterns on top of OFDM transmissions, \proposalName-transmitter~(TX) performs 2D~(i.e., time and frequency domains) amplitude modulation to impose the message-bearing patterns into the waveform and hence convey CTC data across different OFDM-based technologies.
Our approach is best imagined as a punched card from the early days of computers where digital data is represented by the presence or absence of holes in predefined positions -- Fig.~\ref{fig:overlay_ctc}.
We will demonstrate that such message patterns are easily decodeable when cross-observed at a \mbox{\proposalName-receiver~(RX)} although the receiver adheres to a different OFDM-based technology. 

\smallskip
\noindent\textbf{Challenges:} The key principle behind the design of \proposalName~is to embed the intended power patterns into an OFDM signal without corrupting in-technology (i.e. between nodes of the same technology) transmission and in the presence of technology constraints, e.g. lack of fine-grained power control of OFDM resources in LTE and WiFi.

\smallskip
\noindent \textbf{Contributions:} Our key contributions are three-fold:
\begin{itemize}
    \item We analyze the cross-technology observability (shortly \textit{cross-observability}) of specific power variation patterns embedded into OFDM signals when transmitted and received by heterogeneous wireless technologies.
    \item We introduce~\proposalName, a CTC scheme that encodes data as message-bearing power patterns imposed within the OFDM transmission. Its uniqueness comes from its capability to jointly transmit CTC data with high efficiency and negligible overhead to the underlying in-technology communication.
    \item We demonstrate the feasibility of~\proposalName~for the case of for WiFi and LTE-U/LAA, i.e. we design and implement prototype using SDR and COTS hardware. Our evaluations reveal that it achieves reliable and efficient CTC with a bi-directional data rate up to 84\,kbps without significantly affecting in-technology communication.
\end{itemize}

%
\section{Background}

\noindent\textbf{OFDM:} Orthogonal Frequency Division Multiplexing divides the available spectrum bandwidth $B$ into many small and partially overlapping frequency bands called \textit{subcarriers}.
The subcarrier frequencies are selected in such a way that they are \textit{orthogonal} to one another, i.e. signals on subcarriers do not interfere.
In practice, OFDM is efficiently implemented using Fast Fourier Transform (FFT)~\cite{cooley1965algorithm}.
In an OFDM system with FFT size $N$, each subcarrier has the same width of $B/N$ Hz.  
Each subcarrier can be modulated independently (e.g., QAM). 
After modulation, the sender performs an inverse FFT to convert the frequency domain representation into the time domain which is  sent over the air interface. 
The time needed to transmit these $N$ samples is usually called \textit{the FFT period}, which is equal to $N/B\,$sec.
On the receiver side, the OFDM signal is converted back into frequency domain using FFT and each subcarrier is demodulated.
In a nutshell, an OFDM transmitter~(TX) spreads its transmission on a two-dimensional grid which we will refer to as \textit{OFDM time-frequency grid} hereafter.
Some wireless technologies like LTE or 802.11ax use OFDM as multiplexing technique, i.e. they leverage the possibility of assigning subsets of subcarriers to different users.

\smallskip
\noindent \textbf{WiFi:} In 802.11n, the 20\,MHz channel consists of 64 subcarriers with 312.5\,KHz spacing, however only 56 of these 64 are used for communication, occupying the bandwidth of 17.5\,MHz.
The remaining eight subcarriers~(i.e. three and four guards at both bandwidth edges and one DC component in the middle) are \textit{null-subcarriers} that do not carry any signal.
Moreover, four of those 56 subcarriers, so-called \textit{pilots}, are used for channel state estimation.
They are loaded with pseudo-random pilot symbols and their inviolability is crucial for demodulation of WiFi signal.
The FFT period (3.2$\,\mu$s) together with cyclic prefix constitute WiFi symbol (4$\,\mu$s).
WiFi transmits data as self-contained asynchronous frames which can be independently detected and decoded thanks to the prepended preamble and PLCP header (i.e. control data), respectively.
The maximal WiFi frame duration is bound, e.g. to 5.484\,ms in 802.11n. The transmission power can be set on a per frame basis and is the same for all subcarriers.

\smallskip
\noindent \textbf{LTE:} 
An LTE node transmits over a 20\,MHz channel with the sampling rate of 30.72\,MHz using 2048 OFDM subcarriers (15\,KHz spacing).
However, only 1200 subcarriers are used, hence, the occupied bandwidth is equal to 18\,MHz.
The classical LTE transmits a continuous stream of data which is organized into 10\,ms frames each consisting of 10 sub-frames with a duration of 1\,ms.
A sub-frame is further divided into two 0.5\,ms slots and each slot contains 6 or 7 OFDM symbols depending on the duration of a cyclic prefix.
The time-frequency radio resources are organized as \textit{resource blocks~(RBs)}.
A single RB is equal to one slot in time and 12 subcarriers in frequency.
Hence, there are 100 RBs in the 20\,MHz channel.
The resources are grouped into two main physical channels: the
control channel -- Physical Downlink Control CHannel (PDCCH), and the data channel -- Physical Downlink Shared CHannel (PDSCH).
The PDCCH occupies the first 1 to 3 OFDM symbols in each even slot and carries control information including RB-to-UE (User Equipment) assignments.
Moreover, LTE employs two synchronization signals, i.e. Primary and Secondary Synchronization Signals (PSS/SSS) that are carried in the sub-frames 0 and 5 on 62 central subcarriers.
The PSS/SSS signals contain cell information and are used by UEs to achieve time and frequency synchronization with the eNB.
LTE offers a limited DL power control allowing for small adjustment (i.e. in range of [-6\,dB, +3\,dB]) of TX power for all RBs allocated to a single user by means of setting user-specific power offset ($P_{A}$~parameter)~\cite{3gpp.36.213,3gpp.36.331}.

\smallskip
\noindent \textbf{LTE-U/LAA:} LTE leverages carrier aggregation framework to support utilization of the unlicensed bands as secondary component carriers (CC) in addition to the licensed anchor serving as the primary CC. %
Both LTE versions used in unlicensed carriers, i.e. LTE-U and LTE-LAA~\cite{zhang2018lte}, inherit the described frame structure.
The key difference from the classical LTE is their non-continuous channel access.
While \mbox{LTE-U} periodically (de-)activates its unlicensed CC at coarse time scales ($\approx$20 ms duration) through a duty-cycling approach, LTE-LAA relies on Listen-Before-Talk~(LBT) mechanism and achieves finer timescale channel access (1-10 ms).
We will denote by LTE* these unlicensed LTE variants in the rest of the paper unless there is a need to specify the variant.
The frame structure in unlicensed LTE can be simplified when using the \textit{cross-carrier scheduling} (CCS) feature, that allows an eNB to send RB-to-UE mapping for the unlicensed resources in the licensed CC.
In case of CSS, the unlicensed CC does not contain PDCCH and the PDSCH starts from the very first OFDM symbol.
%
%

\section{Exploiting cross-observability}

Here, we analyze OFDM signal modulation in power domain and demonstrate that it can be used as a basis to create a CTC between heterogenous OFDM-based technologies.

\subsection{Cross-observable Power Modulation}

In most cases, due to the different physical layer parameters, the heterogeneous OFDM-based systems are not able to successfully decode the cross-received signals. 
However, the OFDM-based RX may use its FFT module to estimate the power spectral density (PSD)~\cite{pds_estimation} in the points given by the center frequencies of the used subcarriers. 
The resolution of the PSD estimate is determined by its OFDM grid and is equal to the subcarrier spacing~($\Delta f_{\mathrm{RX}}$) in frequency and to FFT period~($\Delta T_{\mathrm{RX}}$) in time.
If the OFDM TX, during its transmission, is able to modulate the TX power of the radio resources at the granularity of RX's PSD resolution, the RX can detect those power changes.
We state that such power variation performed by TX is \textit{cross-observable} by RX.

\begin{figure}[t!]
	\centering
	\includegraphics[width=0.98\linewidth]{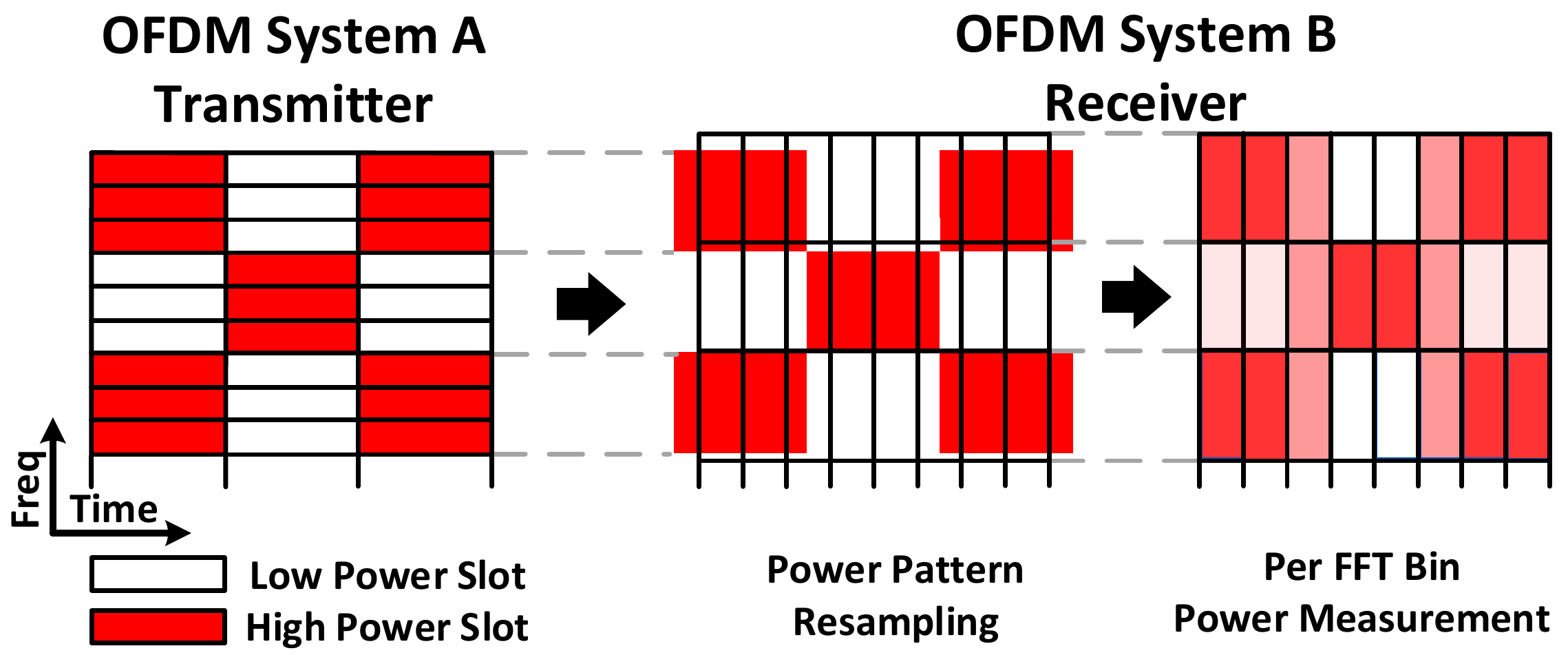}
	\vspace{-10pt}
	\caption{An OFDM TX imposes power patterns into its OFDM time-frequency grid that can be cross-observed, but slightly distorted, at another OFDM RX.}
	\label{fig:ofdmfi_tx_rx}
	\vspace{-15pt}
\end{figure}

Fig.~\ref{fig:ofdmfi_tx_rx} shows an illustrative example, where TX A groups its radio resources into \textit{cross-observable resource blocks}~(CORB) with the size of three subcarriers in frequency and one FFT period in the time dimension and modulates TX power level of each block to create a \textit{cross-observable power pattern}.
The same CORB can be represented (with respect to its bandwidth and duration) in the OFDM grid of RX B with one subcarrier and three FFT periods. Hence, RX B is able to recognize the power pattern imposed by the transmitter in its PSD estimate. 
Although the obtained pattern is recognizable, it is slightly distorted due to asymmetries in the physical layers of both technologies (i.e. sampling rate and FFT size).
This process can be compared to \textit{image resampling}, i.e. transforming a sampled image from one coordinate system to another~\cite{Dodgson92imageresampling}.
Note that we assume silently in this example that TX A is able to control TX power of its radio resources at the highest granularity, i.e. each subcarrier during each FFT period, which is not the case for most of the OFDM-based wireless systems.

The number of the time-frequency resources that have to be grouped together in a CORB depends on the sampling rates and FFT sizes of both involved systems. 
However, according to the Nyquist sampling theorem, reconstruction of the 2D signal is possible when the sampling rate for each dimension is at least twice of the signal bandwidth in the considered dimension.
Hence, given an OFDM-based TX and RX with a different subcarrier spacing $\Delta f$ and symbol duration $\Delta T$, a CORB 
has to satisfy the following equations for its duration $\Delta T_{\text{CORB}}$ and the bandwidth $\Delta f_{\mathrm{CORB}}$:
\begin{equation}
\begin{aligned}
    \Delta T_{\text{CORB}} = a \cdot \Delta T_{\mathrm{TX}} + \epsilon_a \geqslant b \cdot \Delta T_{\mathrm{RX}} + \epsilon_b \label{eq:delta_t}
\end{aligned}
\end{equation}
\vspace{-15pt}
\begin{equation}
\begin{aligned}
    a \geqslant 1,
    b \geqslant 2 \quad \textrm{    where } a,b \in \mathbb{N}, \quad
    \epsilon_a \to 0, \epsilon_b \to 0
\end{aligned}
\end{equation}
\vspace{-15pt}
\begin{equation}
    \Delta f_{\mathrm{CORB}} = n \cdot  \Delta f_{\mathrm{TX}} + \epsilon_n \geqslant m \cdot \Delta f_{\mathrm{RX}} + \epsilon_m  \label{eq:delta_f}
\end{equation}
\vspace{-15pt}
\begin{equation}
\begin{aligned}
    n \geqslant 1,
    m \geqslant 2  \quad \textrm{    where } 
    n,m \in \mathbb{N}, \quad
    \epsilon_n \to 0, \epsilon_m \to 0
\end{aligned}
\end{equation}

\noindent Note that the CORB does not have to be the same in both communication directions. 
Moreover, multiple CORBs may exist, but, higher CTC data rates can be achieved under finer granularity --- see the next subsection. We use $\epsilon$ variables to indicate that small errors in fulfilling the above equations are allowed, i.e. a given value does not have to be necessarily an exact multiple of the corresponding one.
The constraint (4) assures that the power modulation of any CORB (even randomly placed in OFDM grid of TX) can be cross-observed at the RX as long as it is located within RX's bandwidth. However, assuming no carrier frequency drift~(CFD), it can be relaxed to $m \geqslant 1$, when properly placing CORB in OFDM grid of TX, i.e., at the center frequency of one of the RX's subcarriers. 
In such case, the single frequency bin absorbs the entire power of CORB as there is no leakage to adjacent subcarriers~\cite{pds_estimation}.
Our evaluation reveals that in practice small values of CFD of the commodity hardware are tolerable.

The power pattern can be cross-observed only within overlapping spectrum~($\Delta B_{\mathrm{PP}}$) that is computed as the intersection of the spectrum bands of both systems given by intervals:
\begin{align}
\Delta B_{\mathrm{PP}} &= (F^{\mathrm{start}}_{\mathrm{TX}}, F^{\mathrm{end}}_{\mathrm{TX}}) \cap (F^{\mathrm{start}}_{\mathrm{RX}}, F^{\mathrm{end}}_{\mathrm{RX}}).
\end{align}

\subsection{Punched Cards: the Message-bearing Power Patterns}\label{theory}

Having the possibility to exchange cross-observable power patterns, we create the message-bearing patterns and establish a CTC channel for transmission of meaningful data.
Specifically, we use the CORB as a single \textit{CTC-symbol} and modulate its power level. CTC-symbols are organized into a \textit{CTC-grid} with the spacing equal to the duration of one CORB in time (i.e. \textit{CTC-slot}) and its bandwidth in frequency (i.e. \textit{CTC-subcarrier}).
A \textit{CTC-frame} carries data encoded into a message-bearing pattern as depicted in Fig.~\ref{fig:overlay_ctc}.
Note that in the frequency domain, the CTC-frame resembles punched cards, where a hole means bit 0 and its absence bit 1.

Assuming operation at high SNR and large difference between the individual power levels, i.e. allowing to detect them with high probability, the maximum data rate (Nyquist capacity) of the CTC channel equals:

\begin{align}
R_{\mathrm{CTC}} &= \Big\lfloor \frac{\Delta B_{\mathrm{PP}}}{\Delta f_{\mathrm{CORB}}} \Big\rfloor \cdot \frac{\log_2{P}}{\Delta T_{\mathrm{CORB}}} \quad  \left [ \frac{bits}{s} \right ]
\end{align}

\noindent where $P$ is the number of distinguishable transmission power levels. 
Here, we use two power levels, as it is sufficient to create a CTC punched card.
However, our approach can be extended for multiple power levels.

Modulation of all available CTC-symbols may have a negative effect on the underlying in-technology transmissions\footnote{Note that from the perspective of the native technology RX, the modulation of power of already scheduled resources is seen as channel fading.} (i.e. lower data rate and/or a higher bit error rate).
Therefore, we use 1 out of $N$ encoding, i.e. we encode $k$ bits by lowering the power of only one out of $N=2^k$ CTC-subcarriers.
Explaining using our analogy of punched cards, we create a single hole in each column of a punched card.
Furthermore, we can increase the CTC data rate by dividing available CTC-subcarriers into groups and creating a hole in each group.

\subsection{Handling CTC Inter Symbol Interference}\label{PeriodicGridAlignment}
\begin{figure}[t!]
	\vspace{-5pt}
	\centering
	\includegraphics[width=1.0\linewidth]{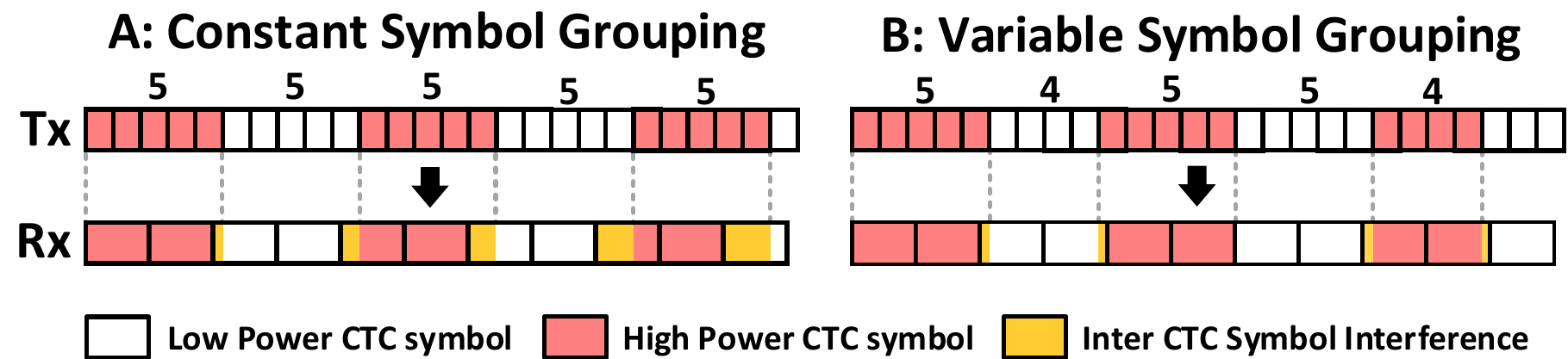}
	\vspace{-15pt}
	\caption{OFDM symbol grouping strategies.}
	\label{fig:ofdmfi_tx_rx_correction}
	\vspace{-15pt}
\end{figure}

In most cases, the duration of the longer symbol (e.g. 71.4$\,\mu$s in LTE) is not exactly a multiple of the duration of the shorter one (e.g. 4$\,\mu$s in WiFi).
The direct effect of this mismatch is inter-symbol interference (ISI) in the time domain, that grows with the number of received CTC-slot ---  Fig.~\ref{fig:ofdmfi_tx_rx_correction}a.
The alignment of the CTC-grid between TX and RX can be periodically corrected by the WiFi node by grouping variable number of OFDM symbols  --- Fig.~\ref{fig:ofdmfi_tx_rx_correction}b.

%

\section{System Design}\label{ofdmfi_design}

Next, we present the design of \proposalName~system supporting the CTC following the concepts from the previous section.

\subsection{\proposalName~Overview}\label{ofdm_ctc}

\begin{figure}[b!] 
	\centering
	\vspace{-15pt}
	\includegraphics[width=0.95\linewidth]{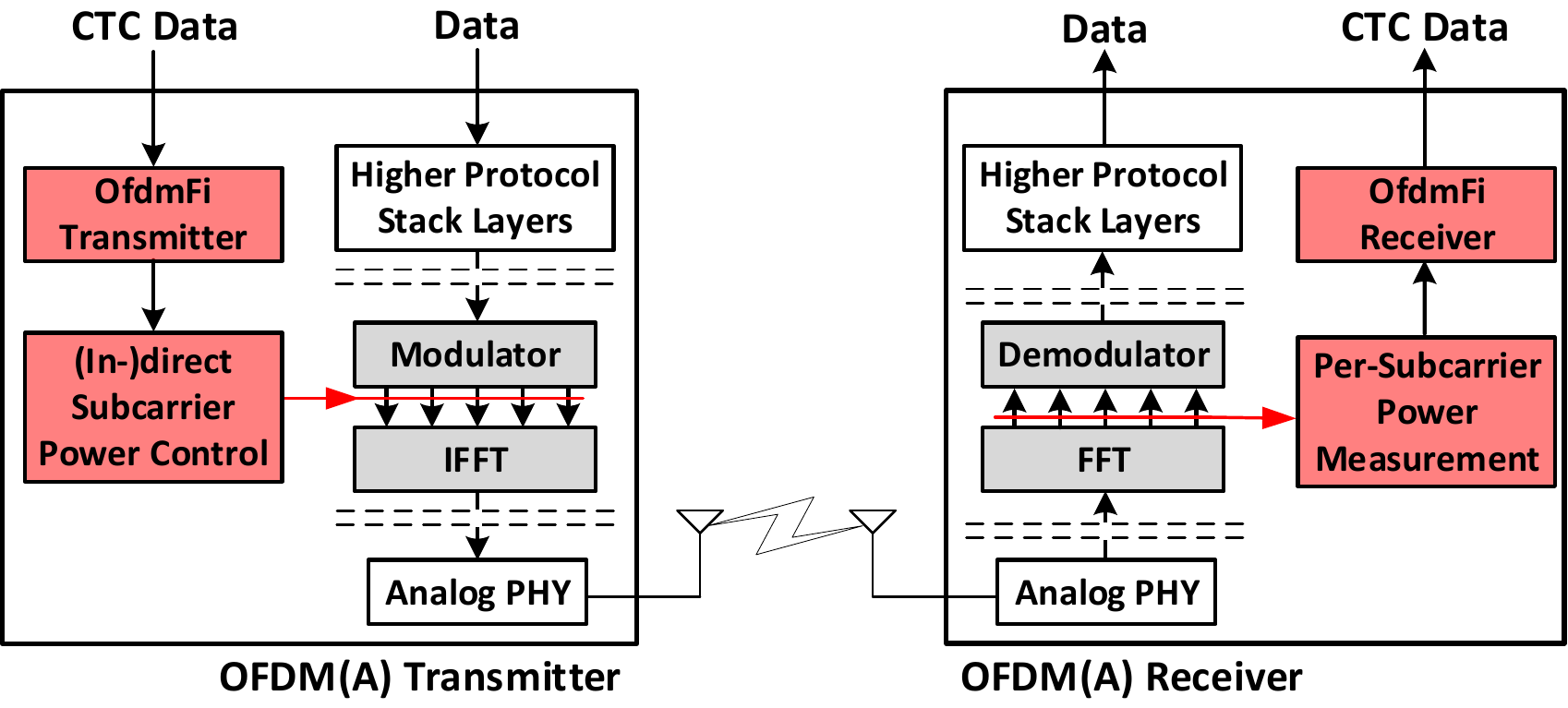}
	\vspace{-8pt}
	\caption{The conceptual architecture of \proposalName~with colored blocks showing the required OFDM technologies' extensions.}
	\label{fig:system_design}
\end{figure}

Fig.~\ref{fig:system_design} shows the conceptual architecture of the \proposalName.
The \proposalName~TX encodes the incoming CTC data and creates a CTC-frame.
The frame is then mapped to a matrix of size matching TX's OFDM grid. The elements are weights between 0 and 1 encoding the intended power pattern.
The matrix is passed to the power control module which applies it row-by-row to the output of the OFDM modulator.
The ability to control the transmission power is essential for the operation of \proposalName.
However, the direct power control is missing in most OFDM-based technologies.
Fortunately, in LTE and WiFi similar effects can be achieved indirectly (see §\ref{practicalOfdmFi}).

At the receiver side, the CTC signal is sampled by performing the power measurements of frequency bins of each OFDM symbol (i.e. the output of the FFT block).
The samples are passed to the \proposalName~RX that aggregates them according to the given CTC-grid and decodes the CTC-frame.
Note that OFDM-RX puts the received signal into its FFT block only after being notified about an incoming transmission (e.g. by frame detection logic in WiFi). Fortunately, the state-of-art commodity WiFi chips (e.g. ath9k and ath10k) offer some limited spectrum sensing capabilities and allow to perform per-subcarrier power measurements with decent rate when the device is not busy with transmission or reception.
We assume similar capabilities in commodity LTE devices and exploit them for our CTC scheme. Note that in 3GPP Release 13, the operation of unlicensed LTE was only specified in downlink (DL), however, the eNBs have to support power sensing capabilities in order to select the least loaded wireless channel in case of LTE-U and enable energy-sensing-based coexistence in case of LTE-LAA. Furthermore, the uplink (UL) operation in the unlicensed channel was included in 3GPP Release 14. Therefore, Rel-14 compliant eNBs are equipped in full RX chain and some parts of it (i.e. FFT block) can be reused for spectrum scanning purposes (like in ath9k or ath10k).

\subsection{Synchronization \& Frame Detection}\label{sec:preabmle}

\begin{figure}[ht!] 
	\centering
	\vspace{-5pt}
	\includegraphics[width=0.75\linewidth]{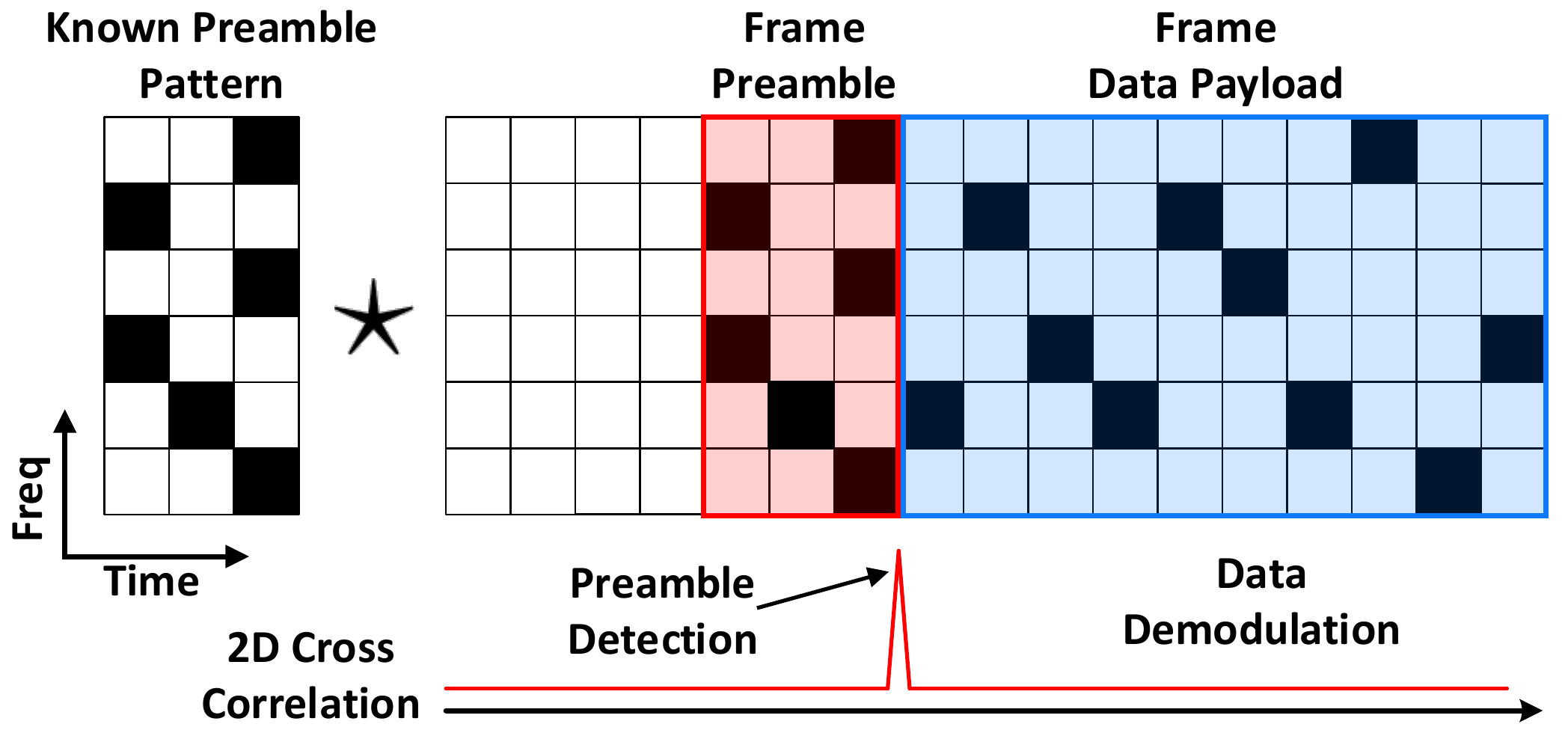}
	\vspace{-10pt}
	\caption{\proposalName~preamble detection mechanism.}
	\label{fig:preamble_detection}
	\vspace{-10pt}
\end{figure}

\proposalName~introduces its own synchronization mechanism based on CTC preamble detection.
Specifically, an \proposalName~TX marks the beginning of a CTC-frame with a predefined preamble, i.e. a unique power pattern, while a receiver is equipped with a preamble detector based on calculating the 2D cross-correlation of the received signal and the known preamble pattern.
The detector computes cross-correlation every time a new row of samples is received from FFT until a peak is detected as illustrated in Fig.\ref{fig:preamble_detection}.
From this point in time, the receiver is synchronized and starts demodulating the CTC-symbols.

\subsection{Channel Estimation \& Demodulation}

An \proposalName~RX follows a classical approach to overcome the channel frequency selectivity, i.e. it performs channel estimation to obtain the reference power level of each CTC-subcarrier.
To this end, it measures their average receive power at the moment of the CTC preamble detection.
The preamble is created using different power levels, however, the changes are known and can be easily reverted.
Afterwards, to reveal the received CTC punched card, the \proposalName~demodulator takes the samples row-by-row and compares the power of each CTC-symbol with the reference level.
If a symbol power level is significantly lower than the reference level, the demodulator marks a hole at its position in the punched card.

\subsection{Channel Access \& Framing}\label{sec:multicell}

\proposalName~only overlays power patterns onto the in-technology frames.
Thus, it completely adopts and depends on the channel access schemes of the underlying technology to avoid collisions and allow efficient multi-node operation.
However, when legacy nodes do not send data frames, \proposalName~TX cannot communicate.
This issue, however, can be solved by introducing a mechanism allowing to trigger transmission of dummy frames in the underlying technology and uses them to carry only valid CTC data.

Finally, to decrease the complexity of the \proposalName~receiver, an entire CTC-frame has to fit into a single continuous transmission attempt of the underlying technology.

%

\section{Punched Cards for WiFi and LTE* CTC}\label{practicalOfdmFi}

On top of the OFDM grid, LTE and WiFi introduce their own logical structures, that cannot be arbitrarily modified.
Therefore, additional constraints have to be considered when integrating \proposalName~with the underlying systems.
To this end, we have to carefully select the subset of radio resources available for CTC.
Our general rule of thumb is to avoid time-frequency resources meant for time and frequency synchronization, channel estimation, and those carrying control data as they are crucial for the proper demodulation of the native signal. Specifically, we impose CTC-frame only on radio resources that carry data.
Moreover, in this section, we address the issue of the missing interface allowing for a fine-grained power control of OFDM resources in both systems.

\subsection{\proposalName~Punch Card Design}\label{wifi_lte_ctc}

\noindent \textbf{WiFi$\rightarrow$LTE*:} The asynchronous nature of WiFi and the lack of fine-grained TX power control of OFDM resources within single transmission prevent implementation of efficient \proposalName-based CTC. However, as we will show in §\ref{card_in_wifi}, it is possible to emulate the missing power control feature at the granularity of a single subcarrier in the frequency and duration of two LTE symbols in time, which allows us to embed CTC-frame within a single WiFi frame. 
Following our rules, we cannot impose CTC pattern in the first part of the WiFi frame, i.e. preamble, PLCP and data header, and cannot use the pilot subcarriers. Therefore, WiFi offers its 52 data subcarriers during data payload for the CTC modulation.

\smallskip
\noindent \textbf{Punched Card:} Out of the 52 available WiFi subcarriers, we take a subset of 48 and divide them into three groups.
In each group, we lower the TX power of a single one in 16 available positions effectively encoding four CTC data bits.
Hence, in a single CTC-slot (i.e. two LTE symbols), we can encode $4 \cdot 3 = 12$\,bits, resulting in a data rate of $ \frac{12\,bit}{2 \cdot 71.4\,\mu s} = 84$\,kbps.

\smallskip
\noindent \textbf{LTE*$\rightarrow$WiFi:} 
We assume the usage of \textit{cross-carrier scheduling} feature in LTE* and exclude the 62 central subcarriers carrying PSS/SSS. Note that they overlap (band-wise) with three WiFi subcarriers, i.e. one null and two data subcarriers assuming operation at the same central frequency.
As we demonstrate in §\ref{card_in_lte}, LTE allows modulating the TX power at the granularity of an RB. Although, the bandwidth of two RBs is slightly wider than that of single WiFi subcarrier, we found that the same punched card with 48 CTC-subcarriers as in case of \mbox{WiFi$\rightarrow$LTE*} can be used.
However, due to longer CTC-slot duration, the expected data rate equals $\frac{12\,bit}{0.5\,ms} = 24$\,kbps.

In Table~\ref{ctc_example_grid}, we summarize the parameters of CORB enabling CTC between WiFi and LTE*. Note that the parameters conform the equations (1-4), e.g. the power modulation of single WiFi subcarrier is observed on roughly 21 LTE subcarriers.

\begin{table}[ht!]
\centering
\vspace{-10pt}
\caption{CORB Parameters for CTC between WiFi (802.11n) and LTE}
\begin{tabular}{c|c|c|c|c|c|c}
\multicolumn{1}{l|}{}   & \multicolumn{3}{c|}{\textbf{Frequency dimension}}              & \multicolumn{3}{c}{\textbf{Time dimension }}                \\
\multicolumn{1}{l|}{}   & \multicolumn{2}{c|}{Subcarriers} & \multicolumn{1}{l|}{}       & \multicolumn{2}{c|}{Symbols} & \multicolumn{1}{l}{}         \\
\textbf{Direction }     & Tx~ & Rx                         & $\Delta f_{\mathrm{CORB}}$  & Tx & Rx                      & $\Delta T_{\mathrm{CORB}}$   \\ 
\hline
WiFi $\rightarrow$ LTE  & 1   & $\approx$21                & 315 KHz                   & 36 & $\approx$2              & 142.8 $\mu$s                 \\
LTE $\rightarrow$ WiFi  & 24  & $\approx$1                 & 312.5 KHz                     & 7  & 125             & 500 $\mu$s                 
\end{tabular}
\label{ctc_example_grid}
\vspace{-10pt}
\end{table}

\begin{figure*}[t]
  \begin{minipage}[b]{0.5\linewidth}
    \centering
	\includegraphics[width=0.8\linewidth]{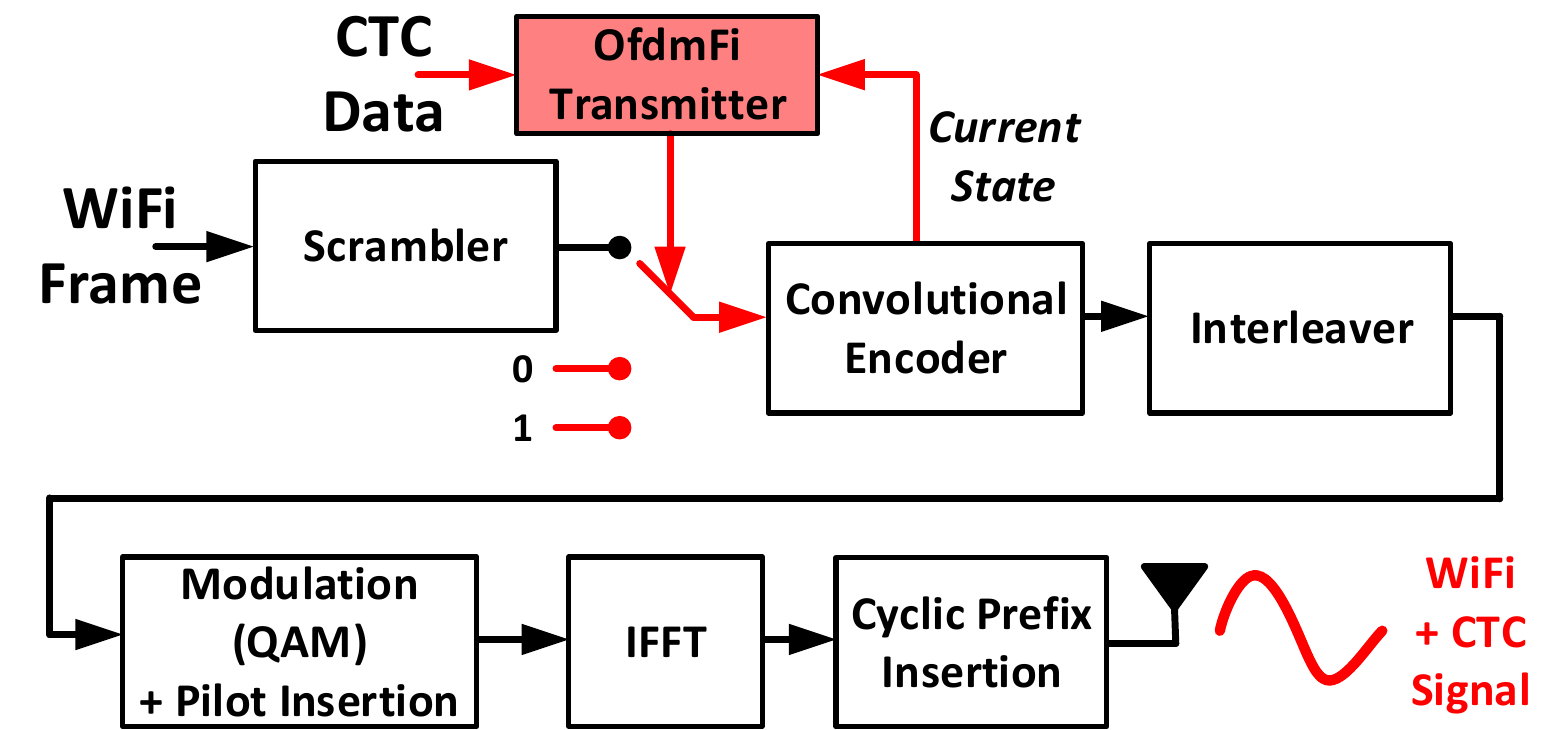}
    \vspace{-10pt}
   \caption{\proposalName~WiFi TX.}
    \label{fig:OfdmFiWiFiTx}
  \end{minipage}\hfill
  \begin{minipage}[b]{0.5\linewidth}
    \centering
    \includegraphics[width=\linewidth]{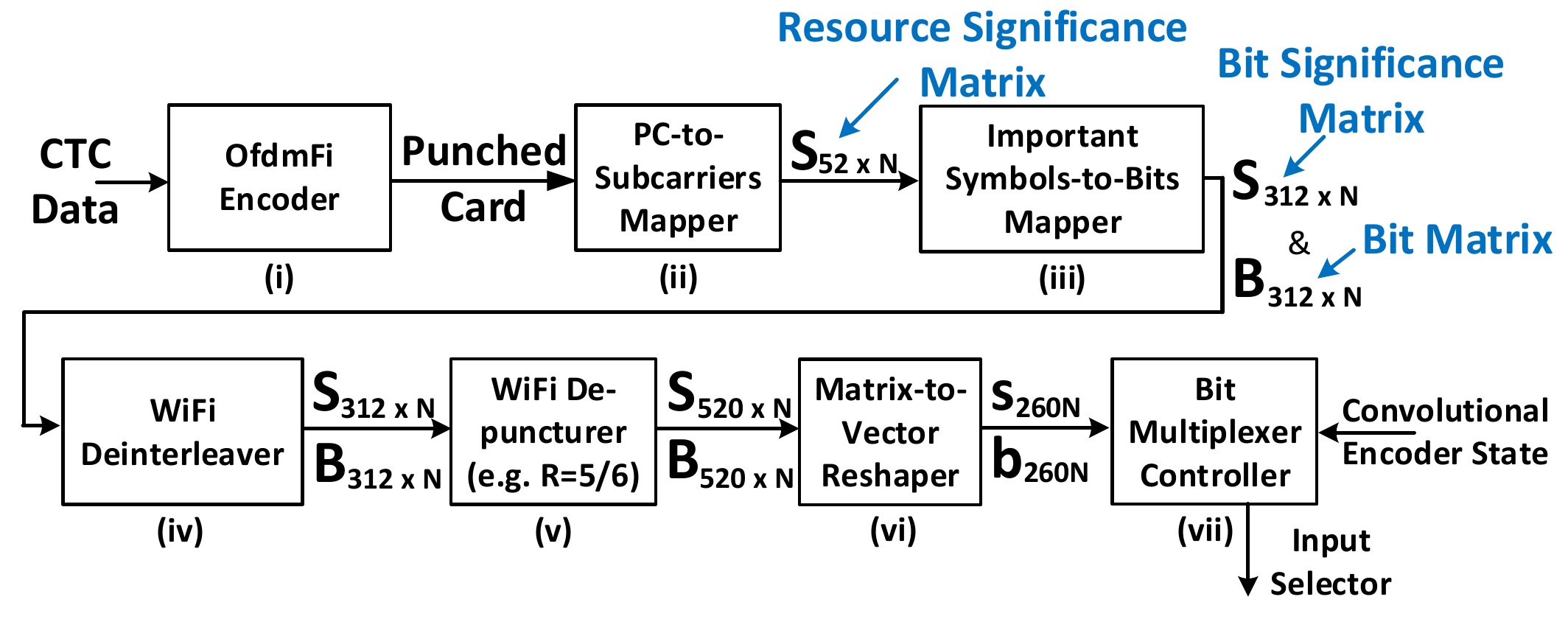}
    \vspace{-20pt}
	\caption{Internal structure of \proposalName~WiFi TX.}
	\label{fig:emulation}
  \end{minipage}\hfill
  \vspace{-15pt}
\end{figure*}

\subsection{Creating CTC Punched Cards in LTE*} \label{card_in_lte}
We envision two approaches to generate punched cards that utilize features provided by LTE standard and require only software updates.
Specifically, we exploit the fact that the Resource Allocation Type 1 (RAT1)~\cite{3gpp.36.213} allows a scheduler to assign the resources at a granularity of a single RB.

\smallskip
\noindent\textbf{RB Blacklisting:} 
The \proposalName~TX blacklists intended RBs in each scheduling round.
Hence, the scheduler omits those RBs when allocating resources.
As the bandwidth of a single WiFi subcarrier corresponds to roughly those of 21 LTE subcarriers, up to 3 RBs overlapping (band-wise) with single WiFi subcarrier have to be blacklisted at the worst case.
Therefore, a throughput drop of up to 9\% (i.e., $3 \cdot 3$ out 100 RBs) is expected for the in-technology communication.

\smallskip
\noindent\textbf{RB Power Control:} 
The \proposalName~TX exploits the built-in DL power control mechanism of LTE. 
Specifically, instead of blacklisting, it just lowers the TX power of particular RBs.
This scheme is beneficial as it creates smaller overhead for the LTE communication as the scheduler may still allocate the RBs with the lower power to UEs experiencing good channel conditions or just slightly decreasing their MCS.

\subsection{Creating CTC Punched Cards in WiFi}\label{card_in_wifi}
\noindent 

To enable commodity WiFi devices to generate punched cards, we propose a standard-compliant per-subcarrier power control emulation through payload modification.
Specifically, we interleave WiFi payload bits with extra bits in the proper positions, i.e. \textit{pattern generating bits} --- Fig.~\ref{fig:OfdmFiWiFiTx}.
Hence, after passing the WiFi TX chain, the modified payload produces the waveform carrying both WiFi payload and CTC power pattern.

\smallskip
\noindent\textbf{TX Power Control Emulation:}
\proposalName~emulates per-subcarrier power control by exploiting the property of M-QAM modulation and the fact that during reception of single CTC-symbol an LTE*-based CTC RX observes only the average power of multiple (i.e. 21) WiFi OFDM symbols.

Specifically, the M-QAM modulation, where $M=2^{b}$, allows encoding groups of $b$ bits into the constellation alphabet given as \mbox{$\alpha_{M-QAM} = \{ \pm(2m-1) \pm (2m-1)j \}$}, where $m \in  \{ 1,...,\sqrt{M}/2 \}$. Therefore, the different constellation points are achieved by modulating both the phase and the amplitude. For instance in 64-QAM, four points close to the center of the constellation diagram ($S_{min}=\pm1\pm1i$) have the smallest amplitude, i.e. $A_{\min}=\sqrt{2}$, while the points in the four corners ($S_{\max} =\pm 7\pm7i$) the highest one, i.e. $A_{\max}=7\sqrt{2}$.
As each point is equally probable, the average power of CTC-symbol observed by LTE*-based \proposalName~RX equals $P_{avg}=A_{avg}^2=42$.

\proposalName~TX forces a WiFi TX to use only the four low amplitude QAM points, i.e. $S_{min}$, on the selected subcarriers for the duration of CTC-symbol. Hence, the averaged symbol power observed by \proposalName~RX equals $P_{\min}=A_{\min}^2=2$. The difference between the average and minimal power levels equals $R_{\textrm{avg}-\min} = 10 \dot \log_{10}(42/2) = 13.22\,\textrm{dB}$. As we will demonstrate, it is sufficient to establish CTC.
Due to space limit, we skip the description of WiFi TX chain and point to~\cite{Perahia}.
To determine when and which bit, 0 or 1, should be interleaved into the WiFi payload, we designed a simple heuristic based on reversing the WiFi TX chain.

In Fig.~\ref{fig:emulation}, we show how the \proposalName~TX creates the pattern generating bits sequence, from step \textit{(i)} to step \textit{(vii)}.
\textit{(i)} The \proposalName~encodes incoming CTC payload and creates a CTC punched card.
\textit{(ii)} The punched card is mapped to WiFi OFDM grid and represented as a resource significance matrix $S_{52,N}$, where 52 refers to the 52 data subcarriers used in 802.11n and $N$ is the number of OFDM symbols.
Specifically, the mapper sets  $S_{ij}=1$ if the subcarrier $i$ during the OFDM symbol $j$ should have low power level~(i.e. be loaded with the low-amplitude 64-QAM constellation point) to encode CTC pattern, otherwise, $S_{ij}=0$.
\textit{%
(iii)} The subcarrier matrix $S_{52,N}$ is transformed to bit matrix $S_{312,N}$ by replicating its columns six times. Now, $S_{ij}=1$ means that this bit is a pattern generating bit and has to be loaded correctly with a proper bit (position-wise) of the low-amplitude constellation point. To this end, a helper matrix $B_{312,N}$ storing required bits on proper positions is constructed. The matrix is filled with the constellation point bits $b_k$ from Table~\ref{min_wifi_symbols} as follows: $B_{i,j} = b_{i\mod6}$ if $S_{i,j}=1$ and is not determined, i.e. $B_{i,j} = x$ otherwise. Note that any bit on positions $b_5$ and $b_2$ (Table~\ref{min_wifi_symbols}) can be used. Hence, they can be removed from the significance bit matrix $S_{312,N}$.
\textit{(iv)} The columns of both matrices $S$ and $B$ are permuted according to the WiFi deinterleaver and \textit{(v)} extended with additional zero bits in the positions determined by WiFi de-puncturing pattern. For instance, the puncturing pattern for the code rate $R=5/6$ is $P_{5/6} = [1, 1, 1, 0, 0, 1, 1, 0, 0, 1]$. Therefore, during de-puncturing, four zero bits are added every six bits which increases size of both matrices from (312,N) to (520,N).
\textit{(vi)} As the convolutional encoder outputs two bits for each incoming bit, both matrices are transformed into vectors of length of 260N storing two bits in each position, i.e. $S_{520,N} \rightarrow s_{260N}$ and $B_{520,N} \rightarrow b_{260N}$.
\textit{(vii)} Having both $s$ and $b$, the bit multiplexer controller knows when to switch between WiFi payload bits (i.e. $s_{j} = b00$) and CTC pattern bits (i.e. $s_{j} \neq b00$). Moreover, in the latter case, it knows what should be the output of the convolutional encoder, i.e. $b_{j}$. However, it does not know yet what should be an input bit to make the encoder generate it.

The convolution encoder used in WiFi can be represented as a finite state machine, where the one input bit activates the transition between states and two output bits are generated during the transition. We observe that all 64 possible states of WiFi encoder can be classified into four groups generating the same output bits when fed with the same input bit --- Table~\ref{conv_encoder_outputs}. Another important observation is that in each state group we can arbitrarily set one of the two output bits by switching the input bit between 0 and 1. 
For instance, when the encoder is in the state from group D, we can put bit 1 to its input to set the next output bit at the position 0 to 0 or put bit 0 as input to set it to 1. Similarly, we can set the output bit at position 1. However, we cannot set both output bits at the same time.

\begin{table}[ht!]
\vspace{-10pt}
\centering
\begin{minipage}[b]{0.65\linewidth}
\centering
\caption{64-QAM Symbols with the Smallest Amplitude}
\label{min_wifi_symbols}
\small
\begin{tabular}{c|cccccc}
\multirow{2}{*}{Symbol} & \multicolumn{6}{c}{Symbol Bits} \\
     & $b_{5}$ & $b_{4}$ & $b_{3}$ & $b_{2}$ & $b_{1}$ & $b_{0}$   \\ 
\hline
18 & {\cellcolor[rgb]{0.753,0.753,0.753}}0 & 1 & 0 & {\cellcolor[rgb]{0.753,0.753,0.753}}0 & 1 & 0   \\
22                         & {\cellcolor[rgb]{0.753,0.753,0.753}}0 & 1        & 0        & {\cellcolor[rgb]{0.753,0.753,0.753}}1 & 1        & 0         \\
50                         & {\cellcolor[rgb]{0.753,0.753,0.753}}1 & 1        & 0        & {\cellcolor[rgb]{0.753,0.753,0.753}}0 & 1        & 0         \\
54                         & {\cellcolor[rgb]{0.753,0.753,0.753}}1 & 1        & 0        & {\cellcolor[rgb]{0.753,0.753,0.753}}1 & 1        & 0         \\ 
\hline
\begin{tabular}[c]{@{}c@{}}Bit\\Mask\end{tabular} & 0 & 1 & 1 & 0 & 1 & 1
\end{tabular}
\end{minipage}\hfill
\begin{minipage}[b]{0.3\linewidth}
\centering
\caption{WiFi Convolutional Encoder State Groups}
\label{conv_encoder_outputs}
\small
\begin{tabular}{c|cc}
\multirow{2}{*}{\begin{tabular}[c]{@{}c@{}}State\\Group \end{tabular}} & \multicolumn{2}{c}{Input Bit}  \\
    & 0  & 1                \\ 
\hline
A & 00 & 11\\
B & 11 & 00\\
C & 10 & 01\\
D & 01 & 10
\end{tabular}
\end{minipage}
\vspace{-5pt}
\end{table}

The \proposalName~TX exploits the above observation to determine the input bit knowing the current state of the convolution encoder and required output $b_{j}$ in the next step. In Table~\ref{ofdmfi_wifi_examples}, we show three examples.

\begin{table}[h!]
\centering
\vspace{-15pt}  
\caption{Example of \proposalName~encoder}
\small
\label{ofdmfi_wifi_examples}
\begin{tabular}{c|c|c|c}
                & Ex.1 & Ex.2 & Ex.3       \\ 
\hline
Importance Mask $s_{j}$& 0\underline{1}   & \underline{1}0    & 00         \\
Required Output $b_{j}$ & x\underline{1}   & \underline{0}x    & xx        \\
Encoder State Group  & C    & A     & x        \\ 
\hline
Input Bit       & 1    & 1     & WiFi Data Bits  \\
Encoder Output  & 0\underline{1}   & \underline{0}0    & xx
\end{tabular}
\vspace{0pt} 
\end{table}

The simple heuristic observes only the current state and does not require any memory. Unfortunately, it may fail when both output bits have to be set, i.e. $s_{j}=b11$. Note that we set one out of two bits arbitrary, while the other one is set correctly with probability of 50\%. However, the matrix $S$ is very sparse, i.e. it contains only 12 significant bits (ones) in each column that are scattered by the WiFi de-interleaver and interleaved with extra zero bits by the WiFi de-puncturer on 520 positions. Although the probability of having two adjacent significant bits is very low, we cannot guarantee to always force the usage of the smallest amplitude constellation points in the required positions of the OFDM grid. However, note that LTE*-based CTC RX observes only the average power of 21 WiFi symbols.
Therefore, even if our approach fails to force low amplitude constellation points in a few out of 21 OFDM symbols (columns) the CTC symbols can be correctly received. Our experiments with interleaving random WiFi with random CTC bits reveal very low fail-rate (i.e. less than 2\%) and confirm the proper operation of \proposalName.
Note that if we force low-amplitude constellation points on only a single WiFi subcarrier in each OFDM symbol, thanks to the de-interleaver, there are no two adjacent significant bits, i.e. $s_{j} \neq b11$. As already mentioned, this property is not necessarily for the operation of \proposalName. Moreover, it comes with a lower CTC data rate, e.g. using 1 out of 32 encoding, we can encode $5$\,bits, resulting in a data rate of $ \frac{5\,bit}{2 \cdot 71.4\,\mu s} = 35$\,kbps.

This approach modifies neither the hardware nor the firmware and introduces only a slight overhead in frame size i.e. 4.6\% in case of MCS 7 as we add 12 every 260 bits\footnote{Note that the control messages for collaboration purposes are not sent with each WiFi frame. Moreover, the expected gains from the collaboration enabled by CTC exceed its overheads.}.
The modified WiFi frame can be received by legacy WiFi receivers, however, some extra steps are required to remove the CTC generating bits from WiFi payload before passing it to the higher layers, i.e. interleaved pattern bits result in a negative result of CRC check, what is an indicator for a STA to remove CTC message and check CRC again.
To this end, we add the length of a CTC message and its bits at the beginning of the WiFi payload. Knowing both the WiFi receiver can generate vector $s$ and use it to puncture the decoded bits in the positions $j$ where $s_{j} \neq b00$. Note that the CTC power pattern cannot be imposed on the WiFi payload part carrying the CTC bits as they have to be decoded first.
%
%

\section{Prototype Implementation}
\begin{figure}
    \vspace{0pt}  
	\centering
	\includegraphics[width=0.79\linewidth]{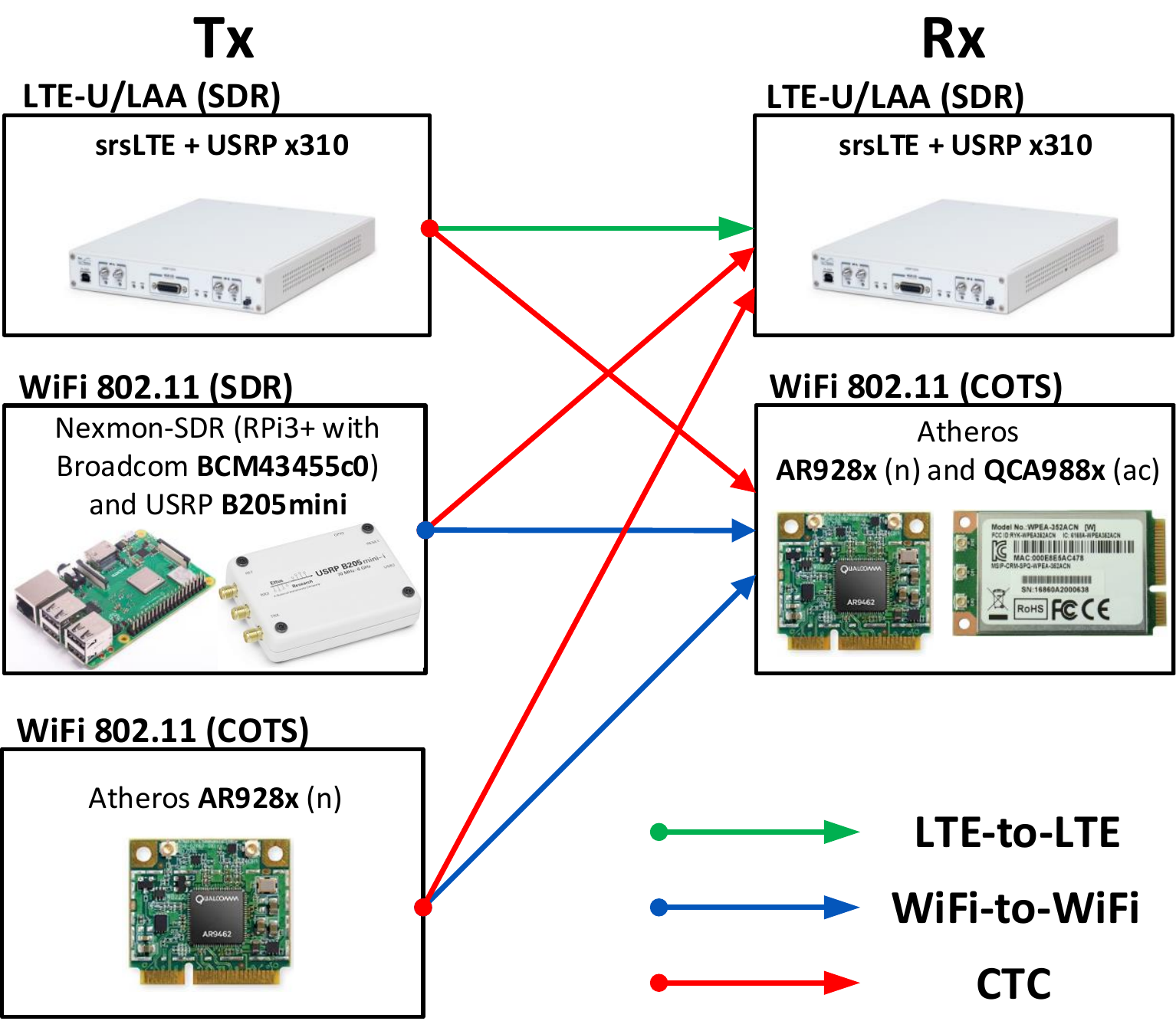}
	\vspace{-5pt}  
	\caption{Platforms used for LTE*-WiFi prototype.}
	\label{fig:ltfi_hardware}
	\vspace{-16pt}
\end{figure}

Next, we introduce the \proposalName~prototype which we have implemented using Software-defined Radio~(SDR) platforms and commodity WiFi devices. 
Fig.~\ref{fig:ltfi_hardware} shows the hardware used for the prototypes.

\subsection{LTE-U/LAA Side}\label{lte_hardware}
To be fully operational~(w.r.t. DL LTE and CTC transmissions), our prototype requires an LTE implementation supporting the following features: \textit{i)} carrier aggregation with licensed and unlicensed carrier components (CC), \textit{ii)} cross-carrier scheduling for the unlicensed CC, \textit{iii)} Resource Allocation Type 1, \textit{iv)} discontinuous channel access in the unlicensed band (duty-cycling or LBT), and \textit{v)} DL power control by means of setting $P_A$ parameter for RBs assigned to a user.

\smallskip
\noindent \textbf{\proposalName~TX:} 
As, to the best of our knowledge, no open-source LTE implementation provides all these features, we selected the srsLTE framework~\cite{srsLte} and introduced direct power control interface in LTE TX, however, sacrificing the operation of the DL channel\footnote{Note that with a more advanced LTE implementation, \proposalName~TX should interface with a MAC scheduler for RB blacklisting and TX power control.}.
Specifically, before performing IFFT, we multiply time/frequency resources of the selected RBs with weights specified by the \proposalName~TX, i.e. in range of [0,1]. 
Since we manipulate the already-scheduled resources, a UE may not be able to receive and decode the PDSCH.
Furthermore, to emulate the discontinuous channel access, we implement a signal gate that is mostly closed, i.e. the time-domain signal is nulled before being passed to the RF front-end.
The duty-cycled access of LTE-U is achieved by opening the signal gate at a slot boundary for a duration of a frame, while LTE-LAA random channel access is emulated by opening the gate at a random point in time.
The \proposalName~TX is implemented in Python and it sends the weights as matrices over TCP socket to the srsLTE transmitter. 
For over-the-air transmission, we use Ettus USRP-X310 SDR platform. 

\smallskip
\noindent \textbf{\proposalName~RX:} We configure LTE node to operate in selected 20\,MHz channel. In addition, we have modified srsLTE to make it always perform FFT operation (i.e. spectrum scanning mode). Then, we copy the output of FFT block, compute the power of each frequency bin and send it to the \proposalName~RX implemented in Python over a socket for CTC decoding.

\subsection{WiFi Side}\label{wifi_prototype}

We selected COTS WiFi devices based on Atheros AR928x (802.11n) and QCA988x (802.11ac) that are controlled with open-source ath9k and ath10k drivers, respectively. 

\smallskip
\noindent \textbf{\proposalName~TX:} 
We generate the content of the WiFi frame carrying both valid WiFi payload and CTC message-bearing pattern using Matlab WLAN Toolbox~\cite{matlabWlan}. Then, we send the frame over a monitor interface~(i.e. raw 802.11 socket). 
Specifically, we have implemented \proposalName~TX as described in §\ref{card_in_wifi}.
First, we interleave the scrambled WiFi payload bits with CTC pattern generating bits.
Second, we pass the modified payload though WiFi de-scrambler as the payload is scrambled again when entering the real WiFi device.
The usage of the same scrambler seed value is essential to generate intended CTC power pattern.
Atheros AR928x chipset increments the scrambling seed value from 1 to 127 (i.e. 7 bits) by one every time it transmits a frame.
In ath5k supported WiFi cards, the scrambling seed can be easily fixed to the value of 1 by setting 0 into the GEN\_SCRAMBLER field in the control register AR5K\_PHY\_CTL (0x992c) of the driver.
Although, not being described, we have found out that the same register and value allow for the same effects also in ath9k WiFi cards.
We have not confirmed this feature in ath10k chipsets. Therefore, in our prototype implementation, the ath9k WiFi cards are used as CTC TX and RX, while ath10k-based chipset only as CTC RX. Note that even without fixing scrambler seed to the value of 1, the CTC message is sent correctly once in 127 transmissions.

Unfortunately, when sending the frames over the monitor interface, the driver does not use the 802.11 MPDU aggregation.
Therefore, the maximal frame size is limited by MSDU size, i.e. 3839 bytes.
As we use MCS 5 (64-QAM, 2/3) during our experiments, the maximal frame duration is bound to $\approx$600\,$\mu s$ and contains only four CTC-slots.
We overcome this issue by generating the waveform of a long WiFi frame (with MPDU aggregation) in Matlab and sending it using USRP SDR.
Moreover, we managed to convert the waveform and successfully send it using the Nexmon-SDR plaform~\cite{nexmon_sdr} based on Raspberry Pi 3+ with Broadcom BCM43455c0 WiFi chip.
However, the platform turned out to be unreliable as only one out of tens of injected frames was transmitted. Therefore, we skip those results.

\smallskip
\noindent \textbf{\proposalName~RX:} Atheros AR928x and QCA988x chipsets provide limited spectral scanning capabilities performing FFT operation and reporting signal strength of each frequency bin $i$, i.e. $|h_i|^2$, at a rate of up to 50\,kHz. 
AR928x performs the 64-point FFT operation for 20\,MHz channel but reports magnitude for 56 subcarriers (i.e. 52 data and 4 pilots).
QCA988x provides up to 256-point FFT for 20\,MHz channel with a resolution of 78.125\,kHz. We use three different configurations for our WiFi-based CTC RX, namely \texttt{Ath9k FFT-64}, \texttt{Ath10k FFT-64} and \texttt{Ath10k FFT-256}.
The samples are delivered only if the device is not busy with TX/RX, hence scanning mode does not affect performance of in-technology transmissions.
The FFT samples are copied from the spectral driver to user space using \textit{relayfs}. 

\begin{figure}[ht!]%
	\centering
	\includegraphics[width=0.95\linewidth]{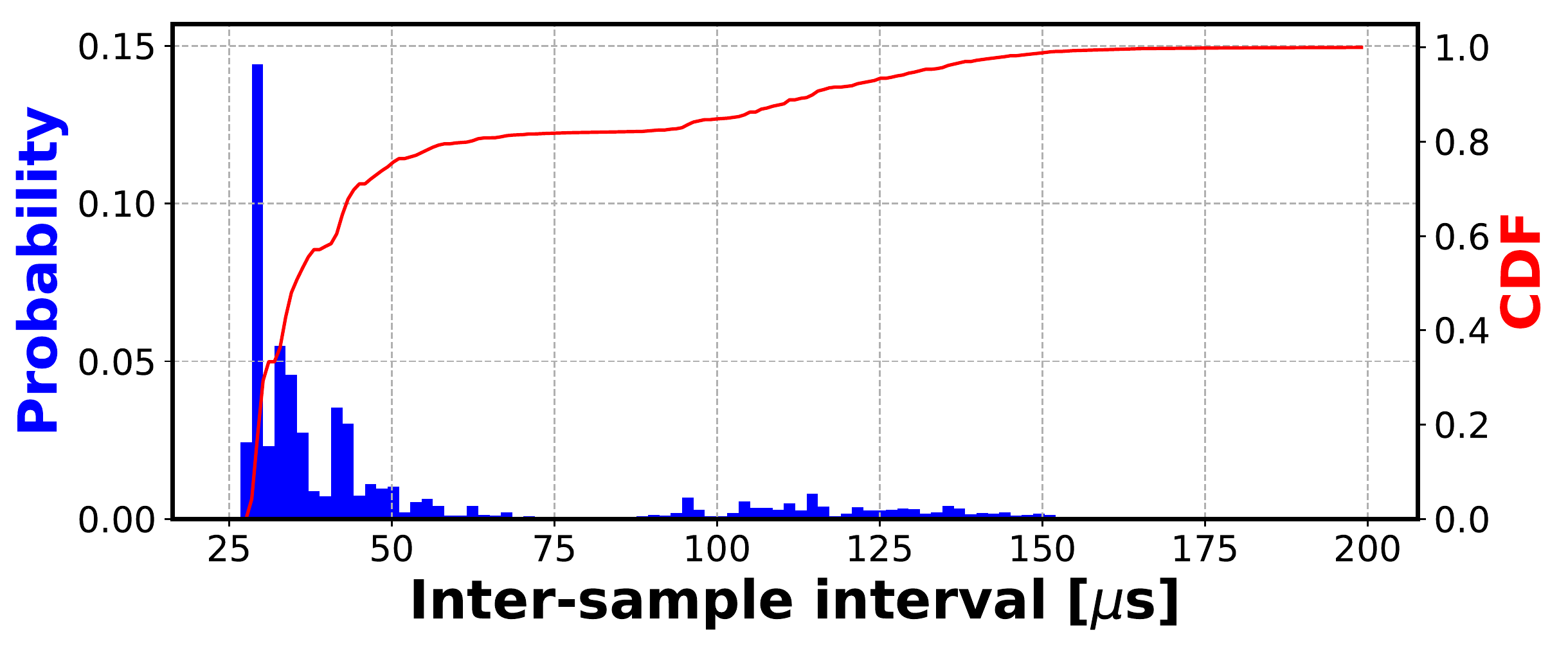}
\vspace{-10pt}  
	\caption{Practical limitations of AR928x WiFi chip. Interval between spectrum samples is not constant.}
	\label{fig:intersample_interval}
	\vspace{-10pt}
\end{figure}

We faced the practical limitation when using those two Atheros chips, namely it turned out that the spectral samples are delivered in irregular periods, i.e. the interval between the majority of FFT samples is lower than $75\,\mu s$, however there are 20\% of the samples which arrive after this value  --- see Fig.~\ref{fig:intersample_interval}.
Fortunately, the samples are time-stamped allowing us for resampling.
Specifically, we collect samples during the $100\,\mu s$ window and aggregate them (i.e. compute the mean value for each frequency bin) before passing to the \proposalName~RX.
If no samples were received during the window, we repeat the last aggregated sample to keep the stream synchronous.
Therefore, we sample a single CTC-symbol (i.e. 0.5\,ms) five times, that conforms to the Nyquist sampling theorem.

%

\begin{figure}[t!]
  \begin{minipage}[b]{0.72\linewidth}
	\includegraphics[width=\linewidth]{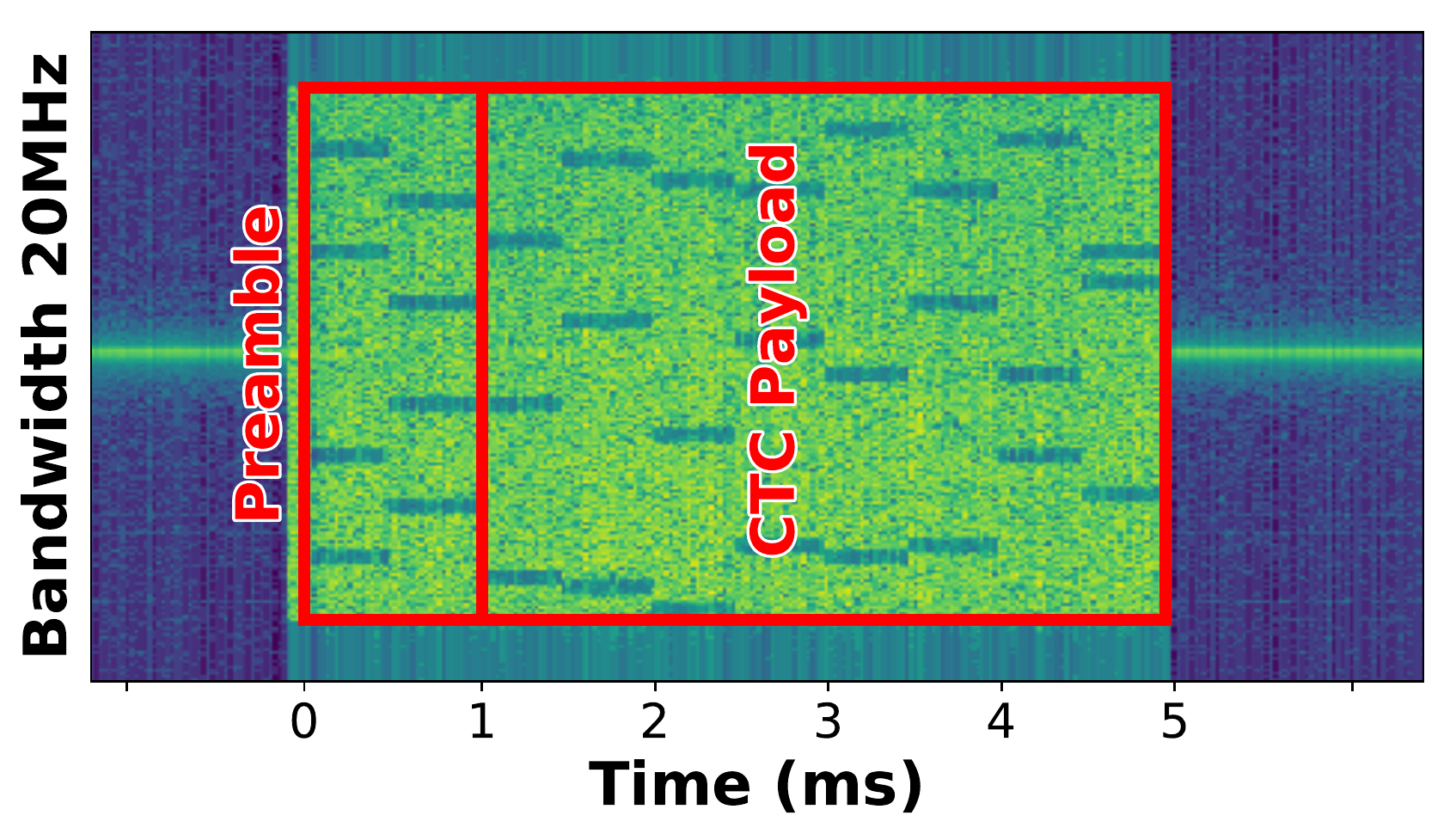}
	\vspace{-20pt}
	\label{fig:ofdmfi_frame_spectogram2} 
  \end{minipage}\hfill
  \begin{minipage}[b]{0.28\linewidth}
    \includegraphics[width=\linewidth]{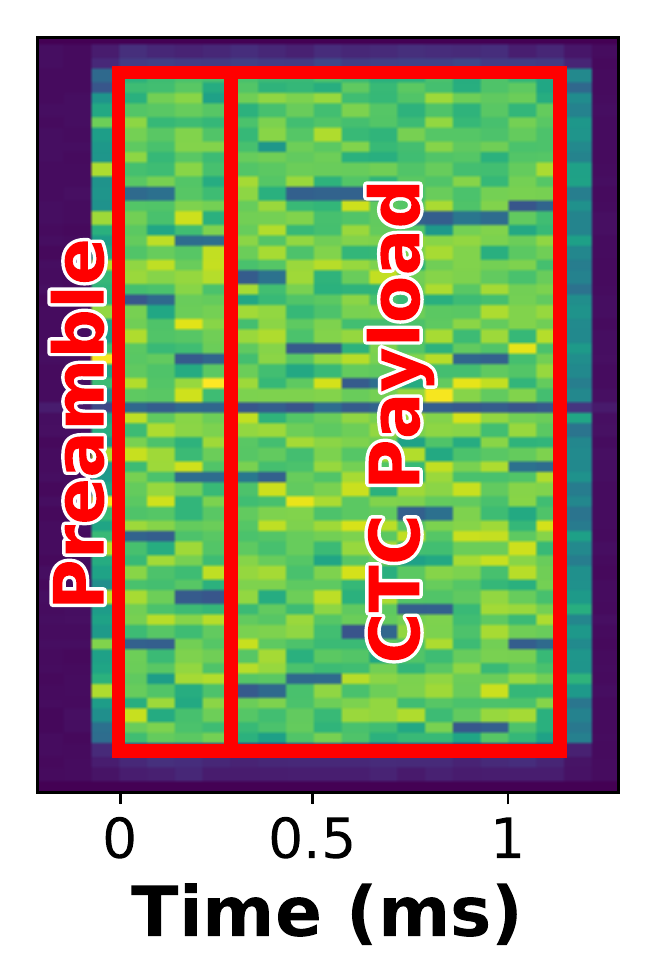}
	\vspace{-20pt}
  \end{minipage}\hfill
  \caption{Spectrograms of punched cards: LTE*$\rightarrow$WiFi and WiFi$\rightarrow$LTE*.	
  \label{fig:ofdmfi_frame_spectogram}}
  \vspace{-15pt}
\end{figure}

\begin{figure*}[t!]
  \begin{minipage}[b]{0.32\linewidth}
	\includegraphics[width=\linewidth]{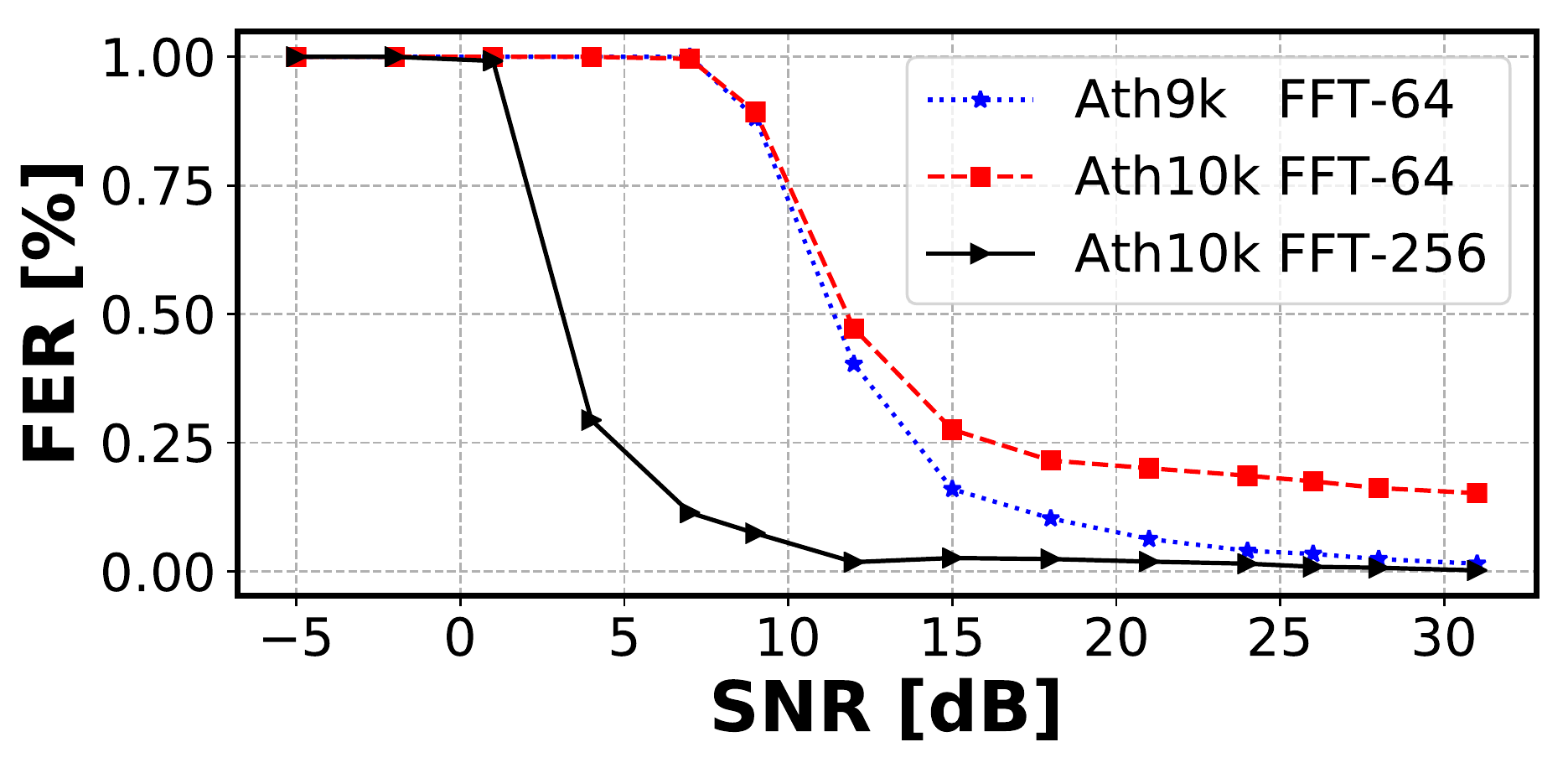}
	\vspace{-20pt}
	\caption{Impact of SNR on LTE*$\rightarrow$WiFi FER.}
	\label{fig:fer_diff_wifichips}
  \end{minipage}\hfill
  \begin{minipage}[b]{0.32\linewidth}
	\includegraphics[width=\linewidth]{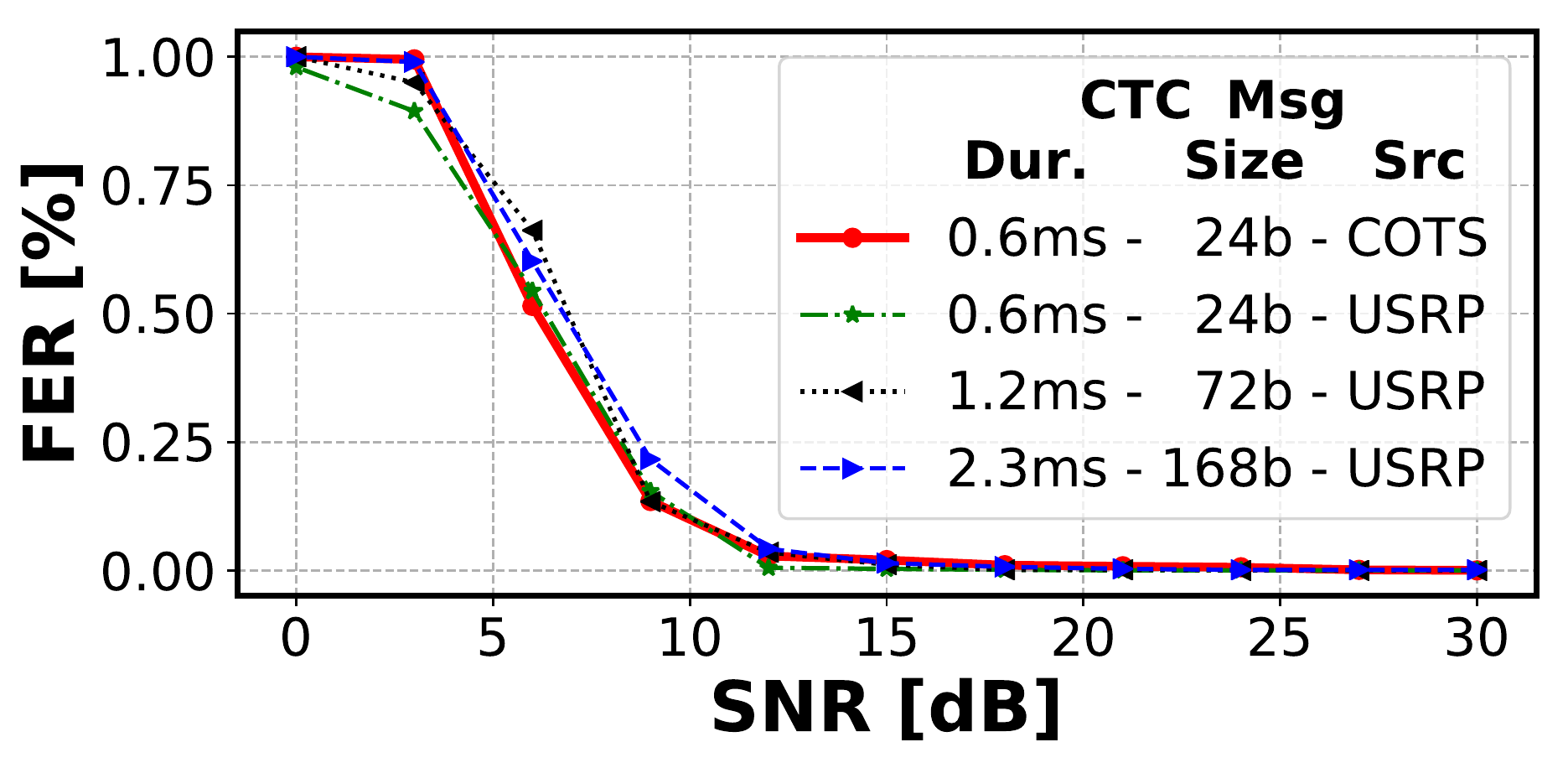}
	\vspace{-20pt}
	\caption{Impact of SNR on WiFi$\rightarrow$LTE* FER.}
	\label{fig:wifi_to_lte}
  \end{minipage}\hfill
  \begin{minipage}[b]{0.32\linewidth}
    \includegraphics[width=\linewidth]{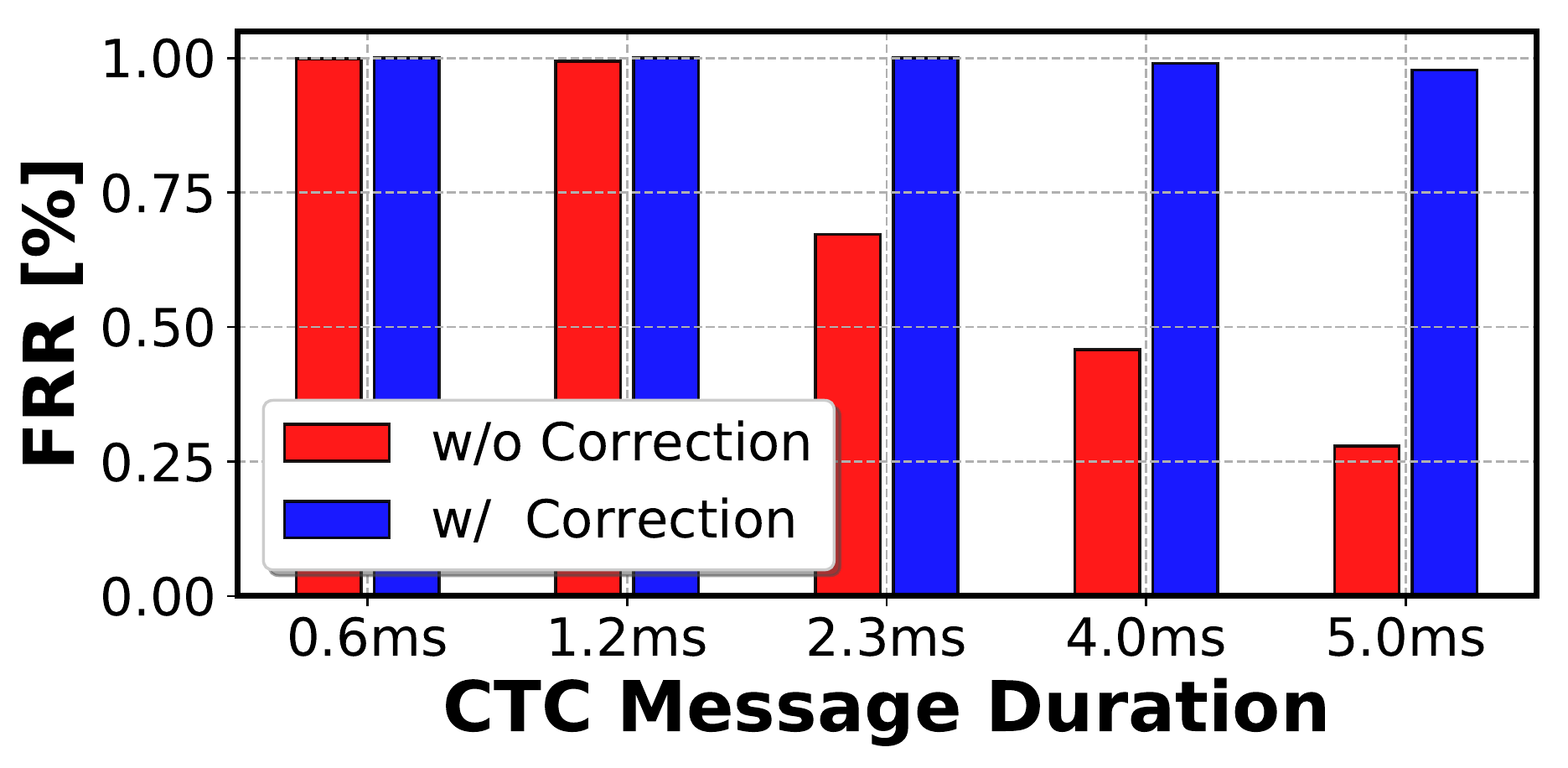}
    \vspace{-20pt}
	\caption{Impact of periodic CTC-slot correction.}
	\label{fig:periodic_grid_correction}
  \end{minipage}\hfill
\end{figure*}

\begin{figure*}[ht!]
  \vspace{-12pt}
  \begin{minipage}[b]{0.32\linewidth}
    \includegraphics[width=\linewidth]{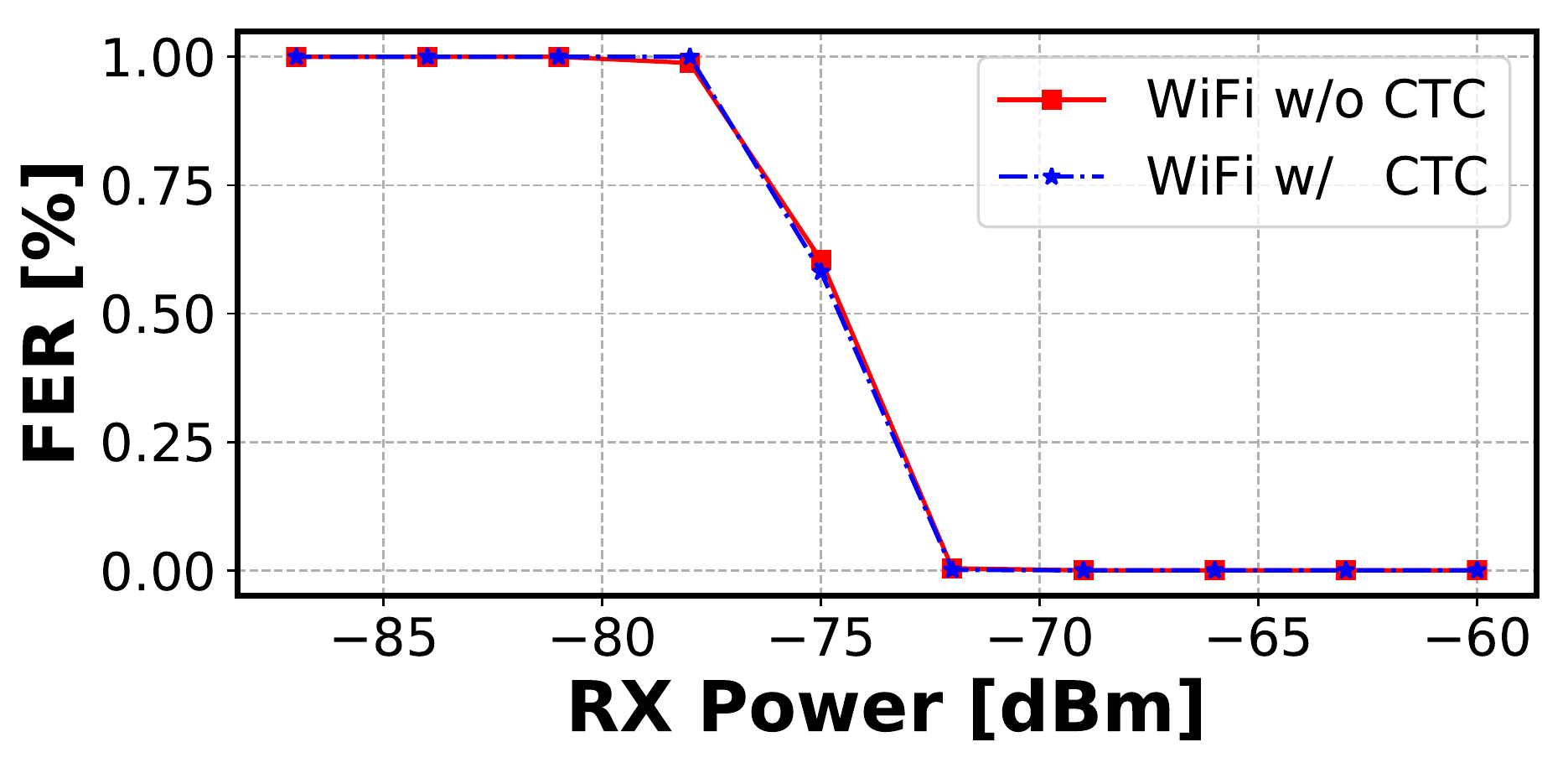}
    \vspace{-20pt}
	\caption{CTC impact on legacy WiFi transmisions.}
	\label{fig:wifi_to_lte_legacy_fer}
  \end{minipage}\hfill
  \begin{minipage}[b]{0.32\linewidth}
    \includegraphics[width=\linewidth]{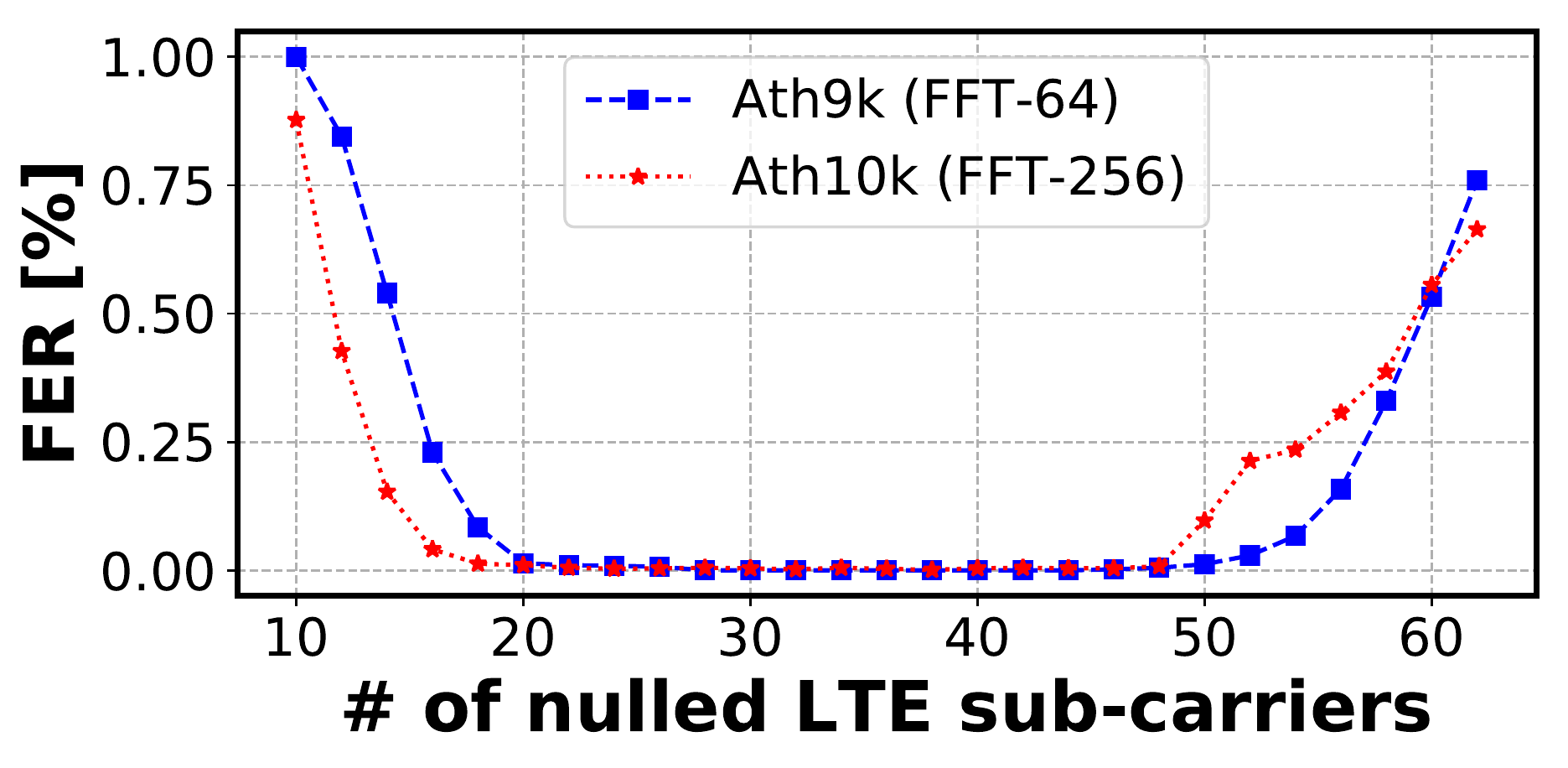}
    \vspace{-20pt}
	\caption{Impact of number of nulled LTE SCs.}
	\label{fig:nulled_sc_num}
  \end{minipage}\hfill
  \begin{minipage}[b]{0.32\linewidth}
	\includegraphics[width=\linewidth]{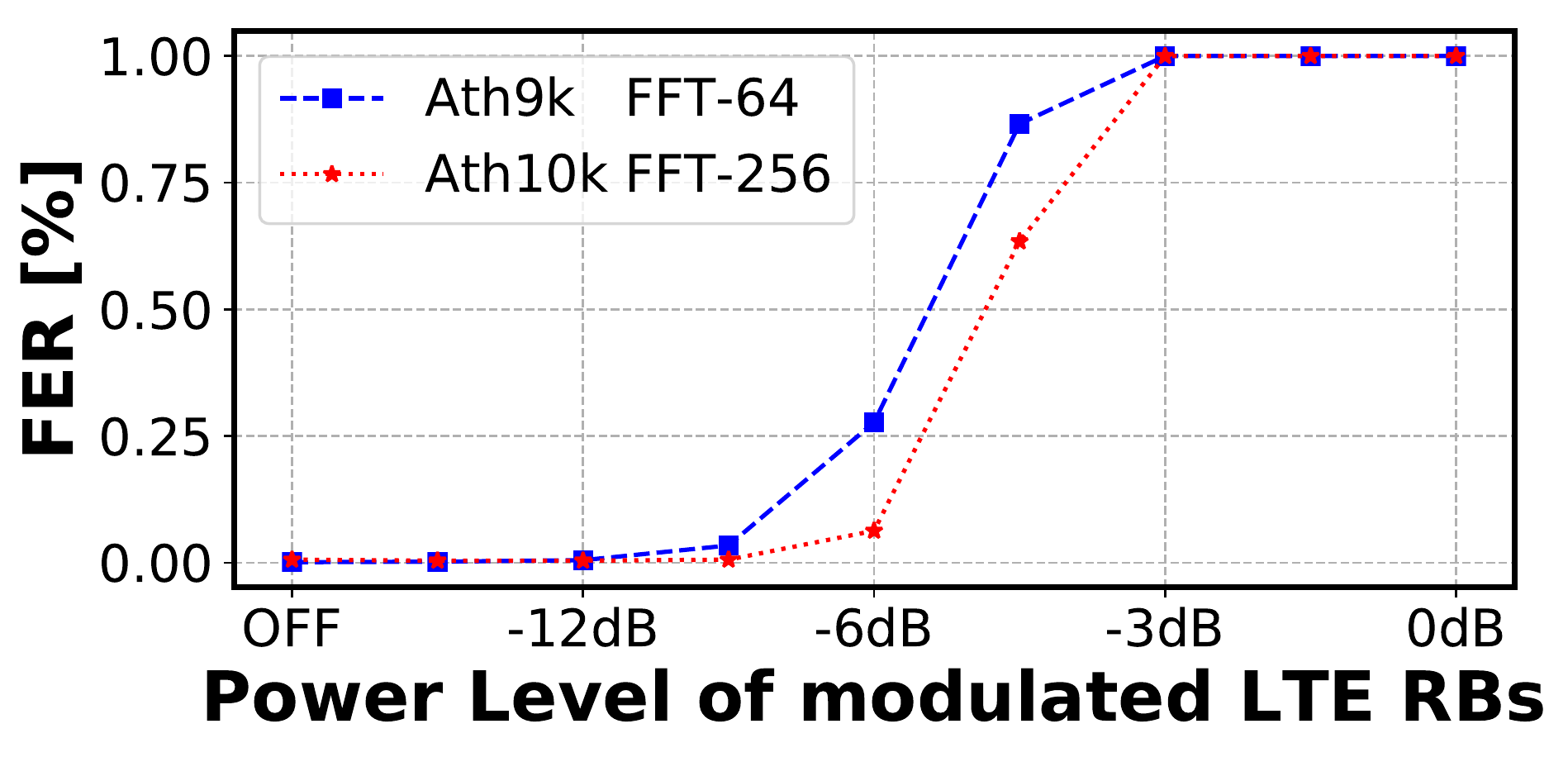}
	\vspace{-19pt}
	\caption{Impact of modulated LTE RBs TX power.}
    \label{fig:pwr_in_modulated_rb}
  \end{minipage}\hfill
 \vspace{-13pt}
\end{figure*}

\section{Performance Evaluation}\label{sec:exp}

We evaluated the performance of the \proposalName~prototype in a small testbed located located in an office space.
During our experiments, all LTE and WiFi nodes operate on the same 20\,MHz channel at 5GHz unlicensed band. The initial distance between nodes was set to 3m.
To ensure statistical significance, each presented result is an average of $10^3$ transmitted frames. 

%
\subsection{Punched Cards over the Air}

The generation of a message-bearing power pattern on top of the legacy in-technology transmission is a base for the CTC.
In Fig.~\ref{fig:ofdmfi_frame_spectogram}, we show the spectrograms of \proposalName~punched cards.
It is easy to recognize the low-power CTC-symbols as well as their duration that equals 0.5\,ms in case of LTE*$\rightarrow$WiFi and 0.142\,ms (i.e. two LTE OFDM symbols) in case of WiFi$\rightarrow$LTE*.
We use three low-power CTC-symbols in each slot to encode data as described in §\ref{wifi_lte_ctc}, while the preamble pattern spans over two CTC-slots and is generated with four low-power CTC-symbols assuring its uniqueness.

\subsection{CTC Frame Error Rate~(FER)}

To evaluate the performance of \proposalName~CTC in terms of frame error rate (FER), we selected an unoccupied wireless channel to avoid external interference and either varied the TX power of the CTC TX or change the distance between nodes to influence the received power and SNR.
The results shown in Fig.~\ref{fig:fer_diff_wifichips} and Fig.~\ref{fig:wifi_to_lte} prove that \proposalName~enables a bidirectional CTC between LTE* and WiFi.
It operates reliably when the SNR exceeds 12\,dB. 

Fig.~\ref{fig:fer_diff_wifichips} presents the FER of LTE*$\rightarrow$WiFi CTC, when using maximal CTC-frame size (i.e. 10\,ms).
The ATH10k-based \proposalName~RX outperforms the ATH9k-based one when using a larger FFT size, i.e. 256 vs. 64.\footnote{In case of 256-point FFT, the power of each CTC-subcarrier is a sum of power values of four FFT bins. This is a known technique to reduce the error of PSD estimate~\cite{pds_estimation}.}
Fig.~\ref{fig:wifi_to_lte} illustrates that there is no difference in FER of WiFi$\rightarrow$LTE* CTC when sending WiFi frame with imposed CTC pattern from COTS and USRP devices.
Moreover, we can see that the increase in the CTC-frame duration has marginal impact on FER.

\subsection{Periodic CTC-Slot Correction}
The duration of CTC-slot has to be periodically corrected to keep it aligned between \proposalName~TX and RX so that errors caused by growing ISI are avoided.
In the case of LTE*$\rightarrow$WiFi CTC, our WiFi-based \proposalName~RX realizes such correction implicitly as it performs resampling of FFT samples to fix the issues related to the variable inter-sample intervals (§\ref{wifi_prototype}).

Fig.~\ref{fig:periodic_grid_correction} illustrates the frame reception ratio (FRR) of CTC between WiFi$\rightarrow$LTE* with and without periodic CTC-slot correction.
During the experiment, the CTC signal was received at the SNR of 30\,dB, hence the degradation of the FRR is attributed to the CTC-slot misalignment.
We observe that the misalignment has a negligible impact for the short CTC-frames, but is harmful for long frames.
Fortunately, using the variable grouping of WiFi OFDM symbols, we can periodically correct the alignment of the CTC-slot between \proposalName~TX and RX and improve the FRR significantly.

\begin{figure*}[t]
  \vspace{0pt}
  \begin{minipage}[b]{0.32\linewidth}
	\includegraphics[width=\linewidth]{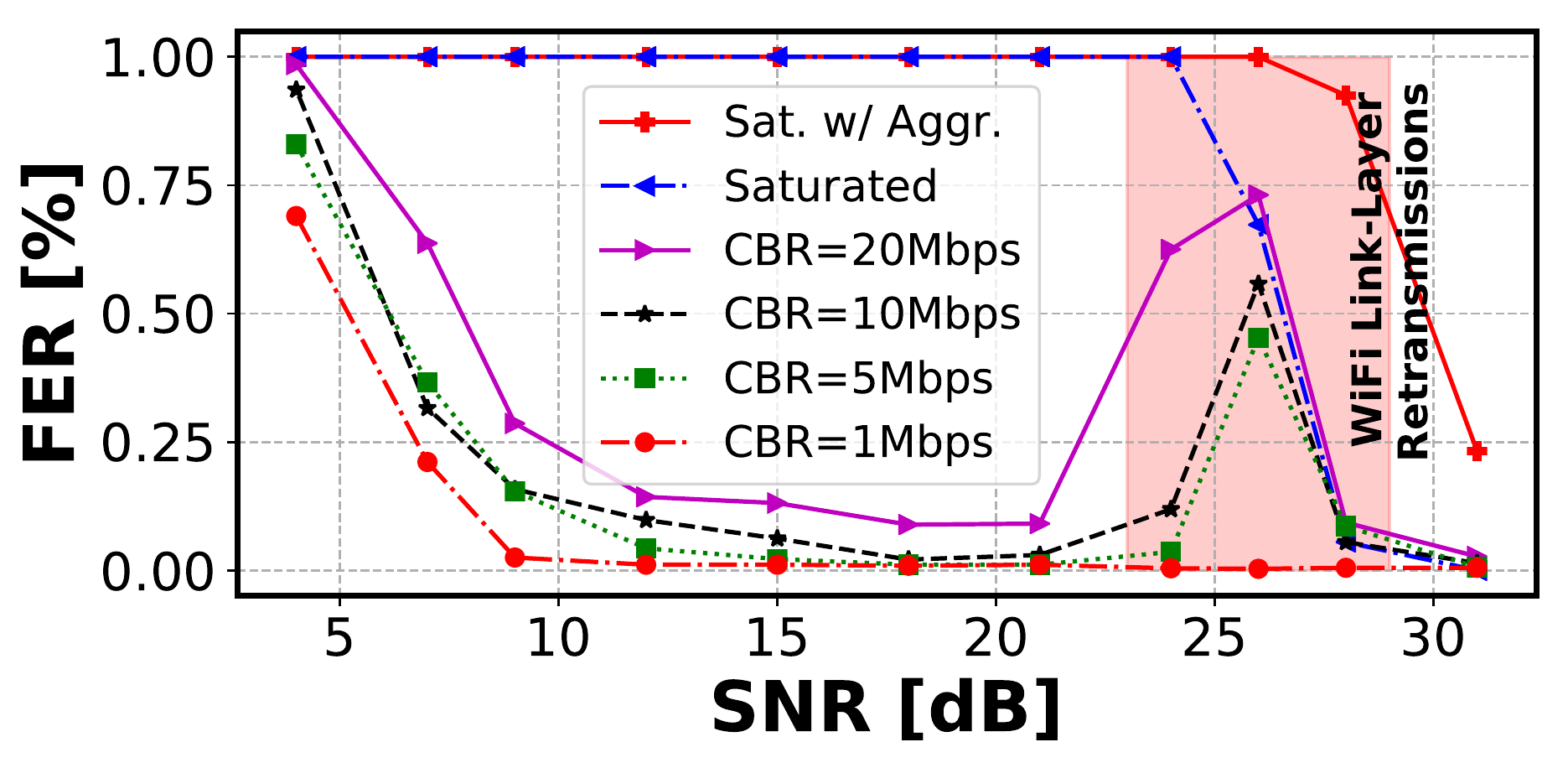}
    \vspace{-20pt}
    \caption{CTC FER under WiFi background traffic.}
    \label{fig:ltfi_ht_fer_bk_ath10k}
  \end{minipage}\hfill
  \begin{minipage}[b]{0.32\linewidth}
    \includegraphics[width=\linewidth]{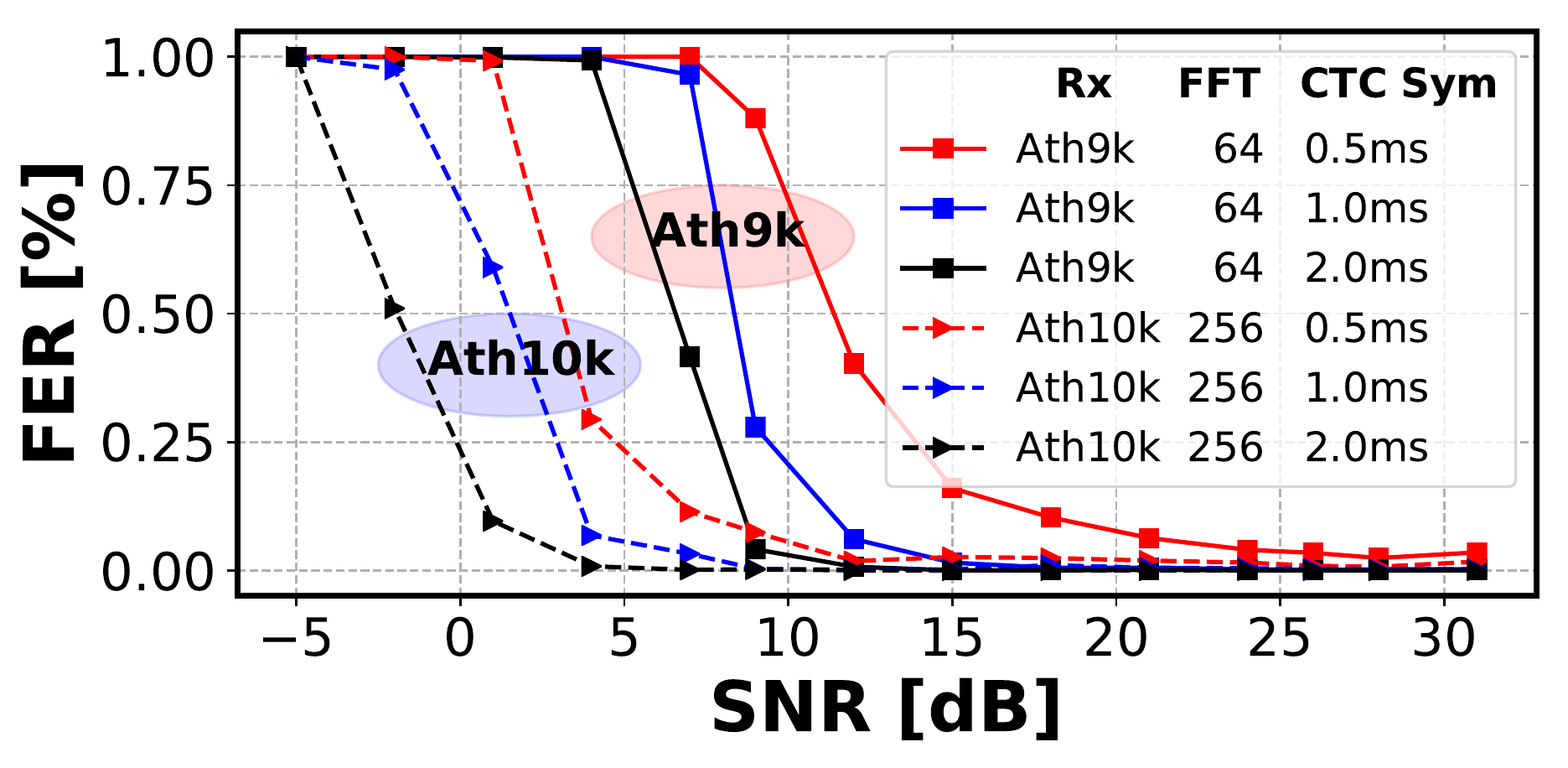}
    \vspace{-20pt}
	\caption{Impact of CTC-symbol duration.}
	\label{fig:lte_to_wifi_ctc_sym_dur}
  \end{minipage}\hfill
  \begin{minipage}[b]{0.32\linewidth}
	\includegraphics[width=\linewidth]{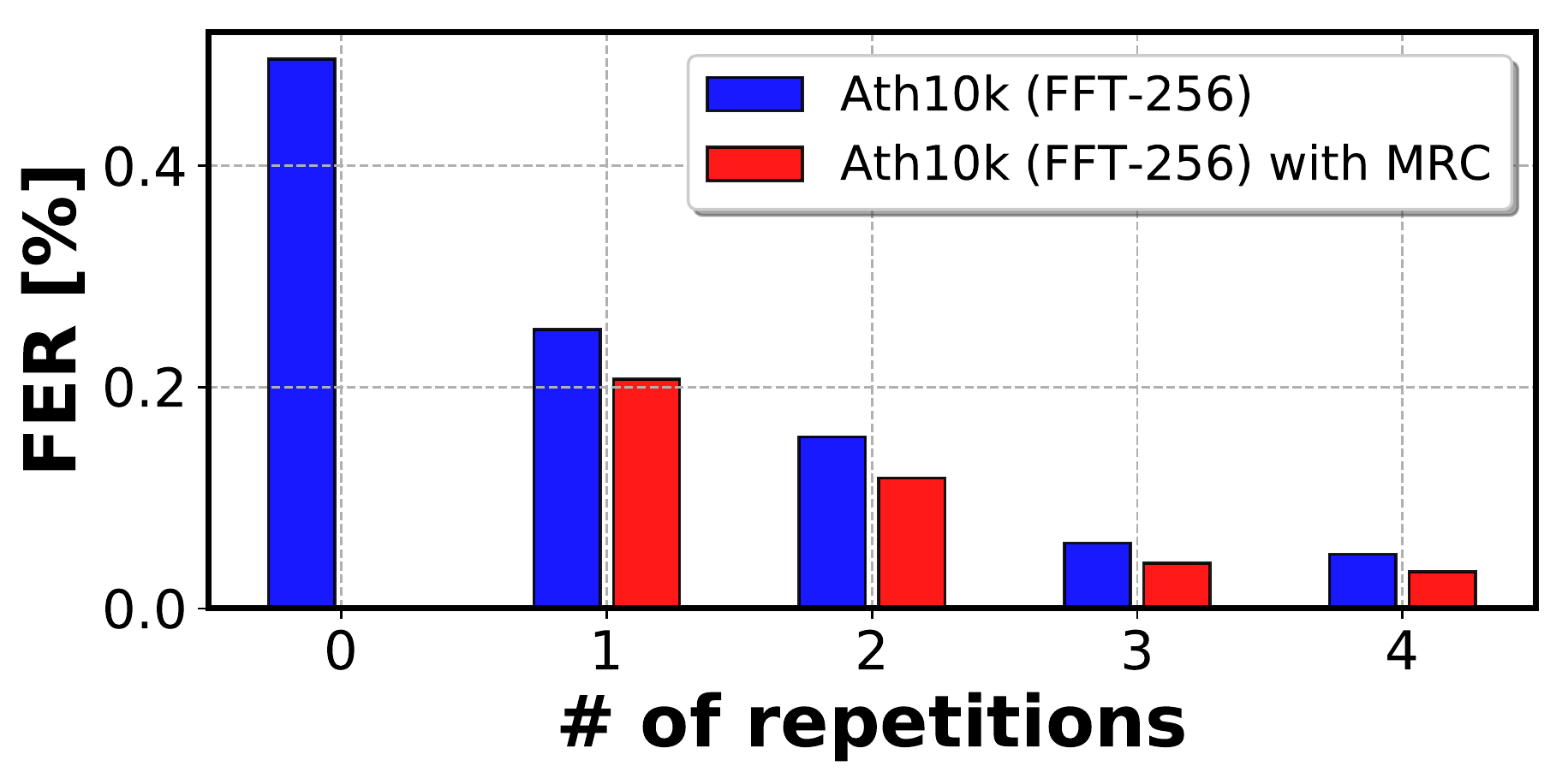}
    \vspace{-22pt}
	\caption{Impact of repetitive transmission.}
	\label{fig:rc_mrc}
  \end{minipage}\hfill
  \vspace{-15pt}
\end{figure*}

\subsection{Impact on In-technology Transmissions}

The WiFi-based \proposalName~TX interleaves WiFi payload bits with extra bits to impose the CTC-frame into the WiFi frame.
To confirm that our approach does not degrade WiFi transmissions, we conduct an experiment, where we first send only standard WiFi frames and then only modified frames (i.e. with CTC) to WiFi RX using MCS-5 (64-QAM, 2/3).
Fig.~\ref{fig:wifi_to_lte_legacy_fer} shows that there is no difference in FER in both cases.
Therefore, the only overhead caused by CTC is the slightly longer frame, e.g. it equals 5.8\,\% in case of MCS-5 as 12 bits are added every OFDM symbol carrying 208 payload bits.

Due to missing features of the used LTE platform (see §\ref{lte_hardware}), we were not able to measure the overhead on LTE caused by our CTC scheme.
However, we can estimate that it does not exceed 9\,\% as the LTE scheduler operates at RBs level and it has to blacklist at most 3 $\times$ 3 out of 100 RBs to impose \proposalName~message-bearing pattern.
However, the bandwidth of the three RBs (36 LTE SCs) exceeds the bandwidth of a single WiFi subcarrier almost two times.
Fig.~\ref{fig:nulled_sc_num} shows the FER of CTC when nulling different number of LTE subcarriers around the center frequency of the corresponding WiFi subcarrier.
First, we observe that \proposalName~operates correctly when nulling between 20 and 48 subcarriers. 
This proves that reliable CTC can be achieved at the level of LTE scheduler and at the granularity of RBs.
Second, for an Ath10k-based \proposalName~RX, nulling of only 16 LTE SCs is enough for a decent CTC operation.
Thus, the overhead of CTC on LTE can be reduced as in most cases the blacklisting of two RBs overlapping (band-wise) with one WiFi subcarrier is sufficient.

Fig.~\ref{fig:pwr_in_modulated_rb} proves that reducing the power level by 9\,dB instead of blacklisting the RBs (OFF) allows a reliable CTC transmission. Note that in LTE standard the maximal TX power reduction is -6\,dB, what allows for decent CTC operation in case of ath10k-baser CTC RX.
Hence, the RBs can be used for LTE* transmissions, which further limits the CTC overheads. 

\subsection{Robustness to Interference}

Since the LTE-U accesses the wireless channel in a duty-cycled manner, the collisions with the WiFi frames are expected. Therefore, we study the impact of the background WiFi traffic on the performance of the LTE-U$\rightarrow$WiFi CTC. 
To this end, in addition to the CTC TX and RX, we placed two additional WiFi nodes in their vicinity and setup a UDP connection with a variable data rate.

Fig.~\ref{fig:ltfi_ht_fer_bk_ath10k} shows the CTC FER performance reported by the Ath10k-based \proposalName~receiver under increasing SNR of the LTE signal. We observe that FER increases with increasing background WiFi traffic. However, the performance drop is bearable (less than 20\%) even with the data traffic at the constant bit rate of 20\,Mbps. In case of saturated data traffic without (i.e. throughput around 32\,Mbps) and with (i.e. 38\,Mbps) the MAC frame aggregation, the \proposalName~receiver can correctly receive CTC-frames only when SNR exceeds 30\,dB (i.e. the LTE-U power is high enough to block WiFi through its energy detection (ED) mechanism). 
Moreover, we can identify the transition area (SNR of 24\,dB and 28\,dB), where the LTE-U signal is not detected by the WiFi TX but is strong enough to corrupt the frames at the WiFi RX. Hence, the frames are often retransmitted leading to higher channel utilization that negatively impacts CTC performance.
\subsection{Increasing CTC Reliability}

We show that \proposalName~can achieve reliable operation when SNR exceeds 12\,dB.
To allow for operation in lower SNR regime, we have two options: i) increase the CTC-symbol duration or ii) transmit the frame multiple times.
Note that both solutions improve the reliability but at the expense of CTC data rate.
Fig.~\ref{fig:lte_to_wifi_ctc_sym_dur} shows the FER of the CTC under different CTC-symbol durations reported by two WiFi-based \proposalName~RXs.
We can clearly see that longer CTC-symbols facilitate operation at lower SNR regime, e.g. with symbol duration of 2\,ms (i.e. four LTE slots) the ath10k-based \proposalName~RX can operate reliably at SNR of 5\,dB. 

Fig.~\ref{fig:rc_mrc} shows the impact of number of retransmissions on the performance of LTE*$\rightarrow$WiFi CTC.
In this scenario, the ath10k-based \proposalName~RX operated at the SNR of 3\,dB and the short CTC-symbol duration was used.
The FER drops with increasing number of retransmissions, e.g. four retransmissions allow for significant FER reduction, roughly by a factor of 10. 
Moreover, Maximum Ratio Combining (MRC) brings further gains in terms of decreased FER by combining the energy from multiple copies of the same frame before decoding.

%

\section{Related Work} \label{chapter:related_work}

The known CTC solutions can be categorized into two classes: i) packet-level CTC and ii) physical-layer CTC.
Approaches belonging to the first class convey the CTC message by modulating bits into either frame length~\cite{Esense,HoWiES}, gap or inter-frame spacing~\cite{TransparentCTC}, packet transmission power~\cite{wizig}, timing of periodic beacon interval~\cite{kim2017free}, prepending legacy packets with a customized preamble containing sequences of energy pulse~\cite{GapSense}, etc.~\cite{cmorse,TransparentCTC,emf}.
\proposalName~can be seen as a generalization of WiZig that modulates the TX power over the whole channel in CORBs spanning over multiple WiFi frames.
The second class, which enables CTC between WiFi and ZigBee, includes WeBee~\cite{WeBee}, TwinBee~\cite{chen2018twinbee} and LongBee~\cite{li2018longbee}.
In WeBee, a WiFi device emulates the ZigBee OQPSK signal by properly selecting payload of WiFi frame.
In \proposalName, we use a similar technique in case of WiFi$\rightarrow$LTE to emulate per subcarrier power control required to create the punched cards.
In contrast to WeBee, our approach is reliable as the punched card signal is much easier to emulate than some real waveform, i.e. frame reception rate close to 100\% vs. only 40-60\% in WeBee. Moreover, in \proposalName~both the in-technology and CTC messages are sent simultaneously at the same time and in the same transmission attempt, while in WeBee the transmission attempt (i.e. WiFi frame) is dedicated to carry either WiFi or CTC bits. 
TwinBee~\cite{chen2018twinbee} improves WeBee in terms of reliability, however, the CTC remains unidirectional.
LtFi~\cite{gawlowicz18_infocom} enables unidirectional CTC from LTE-U to WiFi and promises theoretical data rate up to 665\,bps, however, the presented prototype operates with a rate of 100\,bps. In contrast our \proposalName~enables bidirectional CTC for LTE* and WiFi and offers an increase in data rate by a factor of 125$\times$.
Finally, there are CTC approaches which are not generic as they only target a specific application, e.g. ULTRON~\cite{chai2016lte} for cross-technology virtual channel reservation between LTE* and WiFi.
%

\section{Conclusions}
We propose \proposalName, a CTC scheme enabling direct communication that aims efficient cross-technology collaboration between WiFi and LTE* in unlicensed spectrum.
Using standard-compliant mechanisms, \proposalName~imposes cross-observable data-bearing patterns on top of OFDM transmissions of underlying technologies that can be cross-observed and decoded by heterogeneous receiver.
Our extensive experiments on COTS and SDR platforms revealed that \proposalName~achieves reliable communication with bit rates of 84\,Kbps, (i.e. more than 125x faster than state-of-the-art) while having a marginal impact on the transmissions of underlying technologies.
For the future work, we plan to use our \proposalName~scheme to enable CTC among other OFDM-based wireless technologies.

\newpage

\bibliographystyle{IEEEtran}

{
\footnotesize
\bibliography{biblio,IEEEabrv}
}
\end{document}